\documentclass[]{aastex61}
\usepackage{epstopdf}

\newcommand\aastex{AAS\TeX}

\received{xx}
\revised{xx}
\accepted{xx}
\submitjournal{ApJ}

\shorttitle{\aastex\ sample article}
\shortauthors{Cai}

\begin{document}

\title{Upward Overshooting in Turbulent Compressible Convection. I.Effects of the relative stability parameter, the Prandtl number, and the P\'eclet number.}

\correspondingauthor{Tao Cai}
\email{tcai@must.edu.mo, caitao7@mail.sysu.edu.cn}

\author[0000-0003-3431-8570]{Tao Cai}
\affil{State Key Laboratory of Lunar and Planetary Sciences, Macau University of Science and Technology, Macau, People's Republic of China}
\affil{School of Mathematics, Sun Yat-sen University, No. 135 Xingang Xi Road, Guangzhou, 510275, People's Republic of China}



\begin{abstract}
In this paper, we investigate the upward overshooting by three-dimensional numerical simulations. We find that the above convectively stable zone can be partitioned into three layers: the thermal adjustment layer (mixing both entropy and material), the turbulent dissipation layer (mixing material but not entropy), and the thermal dissipation layer (mixing neither entropy nor material). The turbulent dissipation layer is separated from the thermal adjustment layer and the thermal dissipation layer by the first and second zero points of the vertical velocity correlation. The simulation results are in good agreement with the prediction of the one-dimensional turbulent Reynolds stress model. First, the layer structure is similar. Second, the upper boundary of the thermal adjustment layer is close to the peak of the magnitude of the temperature perturbation. Third, the P\'eclet number at the upper boundary of the turbulent dissipation layer is close to 1. In addition, we have studied the scalings of the overshooting distance on the relative stability parameter $S$, the Prandtl number $\rm Pr$, and the P\'eclet number $\rm Pe$. The scaling on $S$ is not unique. The trend is that the overshooting distance decreases with $S$. Fitting on $\rm Pr$ shows that the overshooting distance increases with $\rm Pr$. Fitting on $\rm Pe$ shows that the overshooting distance decreases with $\rm Pe$. Finally, we calculate the ratio of the thickness of the turbulent dissipation layer to that of the thermal adjustment layer. The ratio remains almost constant, with an approximate value of 2.4.
\end{abstract}




\section{Introduction} \label{sec:intro}

In stars, turbulent convection plays an important role on transporting energy and mixing materials. Also important is the overshooting of turbulent motions from the convectively unstable zone to the adjacent stable one. Early theoretical studies on overshooting are based on phenomenological mixing length models \citep{roxburgh1965note, saslaw1965overshooting, shaviv1973convective,cogan1975convective,maeder1975stellar,roxburgh1978convection}. However, \citet{renzini1987some} has criticized that the extent of overshooting distance highly depends on the assumption implied within these models. Adopting a semi-theoretical approach developed from laboratory experiment and Earth's atmosphere, \citet{schmitt1984overshoot} obtained simple scaling relations of overshooting distance in terms of the initial condition and filling factor through numerical simulations. Later, \citet{zahn1991convective} developed an analytic model on describing the subadiabatic penetration in stars. In his model, he separated the subadiabatic penetration region into two layers: a nearly adiabatic penetrative layer and a thermal adjustment overshooting layer. Under this assumption, he found that the subadiabatic extent ($\delta$) depends on the velocity of penetrative plumes ($W$) with a relation of $\delta\propto W^{3/2}$.

Beyond the mixing length theory (MLT), considerable progress has been made on the development of turbulent theories on time-dependent nonlocal Reynolds stress models (RSMs) \citep{xiong1978stochastic,xiong1989radiation,xiong1997nonlocal,kuhfuss1986model,canuto1992turbulent,canuto1993turbulent,canuto1998stellar,canuto2001ocean,li2007testing,li2012k,li2017applications}.
In RSM, overshooting can be naturally considered as the nonlocal effect of turbulent convection. In stellar evolution, the overshooting distance is usually measured by the extent of the mixing materials in the convectively stable zone. A proper choice of indicator on the extent of overshooting distance is crucial for the prediction of stellar evolution tracks on H-R diagram. \citet{deng2008define} discussed on how to define the boundary of overshooting zone with their model. They suggested to choose the place where the convective flux changes sign as the overshooting boundary. With this choice, their model predicts very extensive overshooting distances in the Sun \citep{deng2008define} and the massive stars \citep{xiong1986evolution}. \citet{zhang2012turbulent} investigated the properties of turbulent overshooting through an asymptotic form of their turbulent convection model. Under the assumption that turbulent dissipation is much stronger than thermal dissipation, they partitioned the overshooting region into three parts: a thin efficient turbulent heat transfer layer, a power-law turbulent dissipation layer, and an inefficient thermal dissipation layer. In an asymptotic mixing model, \citet{zhang2013convective} depicted the convective overshooting mixing as diffusive processes. In the overshooting region, the characteristic length scale of mixing is small and the mixing process is weak \citep{zhang2013convective}. Prediction from the simplified diffusive mixing model has shown good agreement with the fully turbulent convection model \citep{zhang2016simple}. More complete mixing models were developed by \citet{xiong1981statistical} and \citet{canuto2011stellar}, but the direct application of them to stellar evolution is difficult due to the numerical complexity.

The theoretical models mentioned above have made assumptions. The validity of these assumptions needs to be examined. Besides, all these models involve parameters to be determined. These problems could be better understood through numerical simulations. \citet{hurlburt1994penetration} performed a two-dimensional numerical simulation of compressible flow on downward overshooting. They investigated the dependence of subadiabatic extent $\delta$ on stability parameter $S$. They revealed a scaling law of $\delta\propto S^{-1}$ in the penetrative layer and $\delta\propto S^{-1/4}$ in the overshooting layer, respectively. This result is in good agreement with Zahn's analytic model \citep{zahn1991convective}. \citet{freytag1996hydrodynamical} performed two-dimensional numerical simulations with realistic description of radiation and ionization on A-type stars and DA white dwarf stars. They described the overshooting as a diffusive process, and derived an exponential decay parameter for the diffusion. Early attempts of low-resolution three dimensional numerical simulations on overshooting are made by \citet{singh1994three} (upward overshooting), and \citet{singh1995three}, \citet{singh1998study}, \citet{saikia2000examination} (downward overshooting). The scaling laws derived from the numerical simulations of downward overshooting \citep{singh1995three,singh1998study,saikia2000examination} agree well with Zahn's analytical model. Only qualitative result was given in the numerical simulations of upward overshooting \citep{singh1994three}. High-resolution numerical simulations of downward overshooting across a wide range of parameters were presented by \citet{brummell2002penetration}. They confirmed the $-1/4$ scaling law of the thermal adjustment overshooting layer, while the $-1$ scaling law of the nearly adiabatic penetrative layer was absent in the simulations. Based on a semi-analytic model, \citet{rempel2004overshoot} argued that the absence of the nearly adiabatic penetrative layer is caused by the large energy flux specified in the numerical simulations. {Numerical experiments on Boussinesq flow were performed by \citet{korre2019convective}. They reported steeper scaling laws of $\delta\propto S^{-1/3}$ or $\delta\propto S^{-1/2}$, depending on the steepness of the background radiative temperature gradient.} Simulations with realistic physical variables on stellar core convection were performed by \citet{browning2004simulations} and \citet{brun2005simulations}. The effects of rotation and magnetic field are considered. They found that the penetrative convection yields a prolate shape of nearly adiabatic region. \citet{kitiashvili2016dynamics} performed 3D radiative hydrodynamic simulations of the outer layer of a moderate-mass star (1.47 solar mass). Their result discovered a nearly adiabatic layer and a deeper subadiabatic layer. {The recent work of \citet{brun2017differential} simulated the differential rotation and overshooting in solar-like stars. Their result indicated that slow rotators favour a wider overshooting region near the poles and fast rotators at mid-to-low latitude. \citet{hotta2017solar} performed numerical simulations on the solar overshooting with very low energy fluxes $F$. He found that the overshooting distance obeys a scaling law of $\delta\propto F^{0.31}$. \citet{kapyla2018overshooting} conducted numerical experiments on downward overshooting by considering the effect of the smoothness of the heat conduction profiles. He discovered that the power law index of the overshooting distance on the energy flux is smaller in the smooth heat conduction profile than in the step-profile.}  Efforts on prediction of 321D turbulent theory were made by \citet{arnett2015beyond}, \citet{arnett2017synergies}. They separated the overshooting region into three layers: a fully mixed layer, a partially mixed wave layer, and an extra diffusive mixing layer. With the scale analysis of turbulent plumes and eddies , \citet{viallet2015toward} discussed the three possible regimes of turbulent overshooting: diffusion-dominated regime (only mix composition), penetrative regime (transition within the boundary layer), and an entrainment regime (mix both entropy and composition). The selection criterion of different regimes during a stellar evolution calculation is not well defined yet.

In stellar evolution, convective core overshooting is of particular interest because it affects the life time of the stars on the main sequence stage. Using MLT, \citet{claret2017dependence}, \citet{claret2018dependence}, \citet{deheuvels2016measuring} found that the extent of convective core overshooting increases with mass for low-mass stars. \citet{stancliffe2015confronting} calibrated the extent of overshooting by means of eclipsing systems. For most of the eclipsing systems, no single value for overshooting parameters fits them. Asteroseismic analysis of pulsating stars \citep{moravveji2015tight, moravveji2016sub} indicates that exponentially decaying mixing overshooting model \citep{freytag1996hydrodynamical} outperforms those with classic step-function overshoot. Although RSM is difficult to apply, a test of convective core overshooting on binary stars by RSM has shown good agreement with observed data \citep{zhang2012testing}. So far, convective core overshooting is still not well understood. In this paper, we study core convection with 3D numerical simulations of upward overshooting in a Cartesian box.

\section{The Model} \label{sec:model}
The conservation of mass, momentum and energy hydrodynamic equations describing the turbulent compressible convection can be written as
\begin{eqnarray}
\partial_{t}\rho&=&-\mathbf{\nabla}\cdot \mathbf{M} ~,\\
\partial_{t}\mathbf{M}&=&-\mathbf{\nabla}\cdot (\mathbf{MM}/\rho)+\mathbf{\nabla}\cdot \mathbf{\Sigma}-\mathbf{\nabla} p+\rho \mathbf{g}~,\\
\partial_{t}E&=&-\nabla \cdot[(E+p)\mathbf{M}/\rho-\mathbf{M}\cdot\mathbf{\Sigma}/\rho+\mathbf{F_{d}}]+\mathbf{M}\cdot \mathbf{g} ~.
\end{eqnarray}
where $\rho$ is the density, $\mathbf{M}=\rho \mathbf{V}$ is the mass flux, $\mathbf{V}$ is the velocity, $\mathbf{\Sigma}$ is the viscous tensor, $p$ is the pressure, $\mathbf{g}$ is the gravitational acceleration, $E=e+\frac{1}{2}\rho V^2$ is the total energy density, $e$ is the internal energy density, and $\mathbf{F_{d}}$ is the diffusive heat flux.

We use a semi-implicit mixed finite-difference spectral method to solve the `stratified' form of the above hydrodynamic equations \citep{cai2016semi}. The `stratified approximation' assumes the horizontal variations are small compared to the mean values, and the higher order terms containing the products of the horizontal variations are ignored in the momentum and energy equations. Under this approximation, the nonlinear terms are only quadratic and unaliased transform method can be applied efficiently. The derivatives of the prognostic variables are evaluated by Fourier series in the horizontal directions, and by a second-order finite-difference method in the vertical direction. The temporal marching uses a second-order semi-implicit Crank-Nicholson scheme for the linear terms, and a third-order explicit Adams-Bashforth scheme for the nonlinear terms. Since the linear terms are treated implicitly, the advanced time step is not restricted by the CFL conditions arising from acoustic or gravity waves.

To investigate the upward overshooting, we consider the three-dimensional convection with a two-layer setting compressible ideal gas (the ratio of specific heats $\gamma=5/3$) in a cartesian box. The lower layer (layer 1) is convectively unstable, and the upper layer (layer 2) is convectively stable. Initially, the distribution of the gas is assumed in a piecewise linear polytropic state ({the temperature structure is piecewise linear but the heat conductivity is piecewise constant}):
\begin{eqnarray}
T/T_{*} &=& 1+\eta_{i} (1-z)~, \\
\rho/\rho_{*} &=& (T/T_{*})^{m_{i}}~, \\
p/p_{*} &=& (T/T_{*})^{m_{i}+1}~,
\end{eqnarray}
where subscript $*$ represents the value at the interface between these two layers, $0\leq z\leq z_{max}$ is the depth from the bottom, $m_{i}$ is the polytropic index in layer $i$, $\eta_{i}$ is the thickness parameter in layer $i$. All the physical variables can be normalized by the temperature, density, pressure, and height at the interface. After normalization, the temperature, density, and pressure at the interface $z=1$ are ones, and the thermal structures at other locations are integrated upward or downward from the interface. For simplicity of notation, all the physical variables mentioned hereafter are normalized variables. The gravitational acceleration $g=(m_{i}+1)\eta_{i}$ is a constant throughout the whole domain, therefore the flow is in a static state initially. After the definition of \citet{hurlburt1994penetration}, we define the relative convective stability of the two layers by the parameter $S$ :
\begin{equation}
S=\frac{m_{2}-m_{ad}}{m_{ad}-m_{1}}~,
\end{equation}
where  $m_{ad}=1/(\gamma-1)$ is the polytropic index of adiabatic polytrope, and $\gamma=c_{p}/c_{v}$ is the ratio of specific heats, $c_{p}$ is the specific heat capacity {at} constant pressure, $c_{v}$ is the specific heat capacity {at} constant volume. The total flux is
\begin{equation}
F_{tot}=\frac{\kappa_{i}g}{m_{i}+1}~.
\end{equation}
To keep the total flux constant, the thermal conductivity $\kappa_{i}$ is set to be proportional to $m_{i}+1$, therefore the thermal conductivity takes different values at different layers. The ratio of momentum diffusivity to thermal diffusivity is measured by the Prandtl number
\begin{equation}
{\rm{Pr}}(z)=c_{p}\mu/\kappa_{i}~,
\end{equation}
where the dynamic viscosity $\mu$ is a constant throughout the domain. The competition between buoyancy driving and diffusive effects is measured by Rayleigh number
\begin{equation}
{\rm Ra}(z)=\frac{[1-(\gamma-1)m_{i}](m_{i}+1) \eta_{i}^2\rho^2{\rm Pr(z)}}{\gamma \mu^2}~.
\end{equation}
The ratio of inertial forces to viscous forces is measured by Reynolds number
\begin{equation}
{\rm Re}(z)=\frac{\rho v''L_{1z}}{\mu}~,
\end{equation}
where $v''$ is the root mean square value of velocity, and $L_{1z}$ is the height of layer 1. The ratio of convective transport rate to diffusive transport rate is defined by P\'eclet number
\begin{equation}
{\rm Pe(z)}={\rm Re(z)}\frac{c_{p}\mu}{\kappa_{i}}~.
\end{equation}

We use a stress-free boundary condition on the vertical direction, and periodic boundary conditions on the horizontal directions. The temperature is fixed at the top boundary, and a constant flux is supplied at the bottom boundary. The boundary conditions on the vertical directions are expressed as
\begin{eqnarray}
v_{z}=\partial_{z}v_{x}=\partial_{z}v_{y}=0, \quad T=constant, \text{\quad at $z=z_{top}$}  ~,\\
v_{z}=\partial_{z}v_{x}=\partial_{z}v_{y}=0, \quad F_{tot}=constant, \text{\quad at $z=z_{bot}$}~,
\end{eqnarray}
where $v_{x}$, $v_{y}$, and $v_{z}$ are velocities along x-, y-, and z-directions, respectively. The subscripts $top$ and $bot$ represent the top and bottom layers of the simulation box.

We have run a total of 13 cases with varying relative stability parameter $S$, Prandtl number $\rm Pr$ and P\'eclet number $\rm Pe$. Cases A1-A7 vary the parameter $S$ while keeping the $\rm Pr$ and $\rm Pe$ almost constant. Cases A3 and B1-B3 vary $\rm Pr$ while keeping $S$ and $\rm Pe$ constant. Cases A3 and C1-C3 vary $\rm Pe$ while keeping $S$ and $\rm Pr$ constant. For all the simulations, we set $\gamma=5/3$, $m_{1}=1$, $\eta_{1}=2$, and $z_{top}=1.5$. The aspect ratio of the simulation box is $4:4:1$. {The depth of the convectively unstable zone is about 2.2 pressure scale heights. The depth of the convectively unstable zone varies from 2.4 to 3.3 pressure scale heights, depending on the stability parameters (or polytropic indices).} The details of the simulation parameters are listed in Table~\ref{tab:table1}.

\startlongtable
\begin{deluxetable}{cccccccccccc}
\tablecaption{Parameters of numerical simulations \label{tab:table1}}
\tablehead{
 Case & $N_{x}\times N_{y}\times N_{z}$ & $S$ & $\mu$ & $F_{tot}$ & $\rm Ra$ & $\rm Pr$ & $\langle v''\rangle_{cz}$ & $\rm Re$ & $\rm Pe$ & $\delta_{1}$ & $\delta_{2}$
 }
\startdata
 A1   & $512^2\times 301$ & $1$   &   $1.25\times 10^{-4}$    & 0.00125   &   $2.11\times 10^{8}$   & $0.5$    & 0.070  & 1102  & 550.8  & 0.13  & $>0.5$\\
 A2   & $512^2\times 301$ & $2$   &   $1.25\times 10^{-4}$    & 0.00125   &   $2.11\times 10^{8}$   & $0.5$    & 0.068  & 1084  & 542.2  & 0.10  & 0.335\\
 A3   & $512^2\times 301$ & $3$   &   $1.25\times 10^{-4}$    & 0.00125   &   $2.11\times 10^{8}$   & $0.5$    & 0.067  & 1062  & 530.9  & 0.09  & 0.305\\
 A4   & $512^2\times 301$ & $4$   &   $1.25\times 10^{-4}$    & 0.00125   &   $2.11\times 10^{8}$   & $0.5$    & 0.067  & 1062  & 530.8  & 0.08  & 0.285\\
 A5   & $512^2\times 301$ & $5$   &   $1.25\times 10^{-4}$    & 0.00125   &   $2.11\times 10^{8}$   & $0.5$    & 0.067  & 1059  & 529.6  & 0.075 & 0.285\\
 A6   & $512^2\times 301$ & $6$   &   $1.25\times 10^{-4}$    & 0.00125   &   $2.11\times 10^{8}$   & $0.5$    & 0.066  & 1055  & 527.2  & 0.075 & 0.26\\
 A7   & $512^2\times 301$ & $7$   &   $1.25\times 10^{-4}$    & 0.00125   &   $2.11\times 10^{8}$   & $0.5$    & 0.066  & 1034  & 516.9  & 0.075 & 0.255\\
 B1   & $512^2\times 301$ & $3$   &   $2.5\times 10^{-4}$     & 0.00125   &   $1.05\times 10^{8}$   & $1.0$    & 0.064  & 509   & 509.0  & 0.095 & 0.33\\
 B2   & $512^2\times 301$ & $3$   &   $6.25\times 10^{-5}$    & 0.00125   &   $4.22\times 10^{8}$   & $0.25$   & 0.070  & 2213  & 553.3  & 0.085 & 0.28\\
 B3   & $512^2\times 301$ & $3$   &   $3.125\times 10^{-5}$   & 0.00125   &   $8.44\times 10^{8}$   & $0.125$  & 0.071  & 4489  & 561.1  & 0.08  & 0.27\\
 C1   & $512^2\times 301$ & $3$   &   $2.5\times 10^{-4}$     & 0.00250   &   $5.27\times 10^{7}$   & $0.5$    & 0.084  & 669   & 334.5  & 0.12  & $>0.5$\\
 C2   & $512^2\times 301$ & $3$   &   $6.25\times 10^{-5}$    & 0.000625  &   $8.45\times 10^{8}$   & $0.5$    & 0.054  & 1704  & 851.8  & 0.065  & 0.215\\
 C3   & $512^2\times 301$ & $3$   &   $3.125\times 10^{-5}$   & 0.0003125 &   $3.38\times 10^{9}$   & $0.5$    & 0.043  & 2730  & 1364.2  & 0.05 & 0.16\\
 \enddata
\tablecomments{$N_{x},N_{y},N_{z}$ are the numbers of grid points in x-, y-, and z-directions, respectively. $S$ is the relative stability parameter. $\mu$ is the dynamic viscosity. $F_{tot}$ is the total flux. $\rm Ra$ is the averaged Rayleigh number. $\rm Pr$ is the averaged Prandtl number. $\langle v''\rangle_{cz}$ is the averaged root mean square velocity. $\rm Re$ is the averaged Reynolds number. $\rm Pe$ is the averaged P\'eclet number. $\delta_{1}$ is the distance measured by the first zero of velocity correlation. $\delta_{2}$ is the distance measured by the second zero of velocity correlation. All the averaged values are taken both temporally and spatially in the convection zone.}
\end{deluxetable}

\section{Numerical Results}
\subsection{The extent of overshooting distance}
\begin{figure}
\gridline{\fig{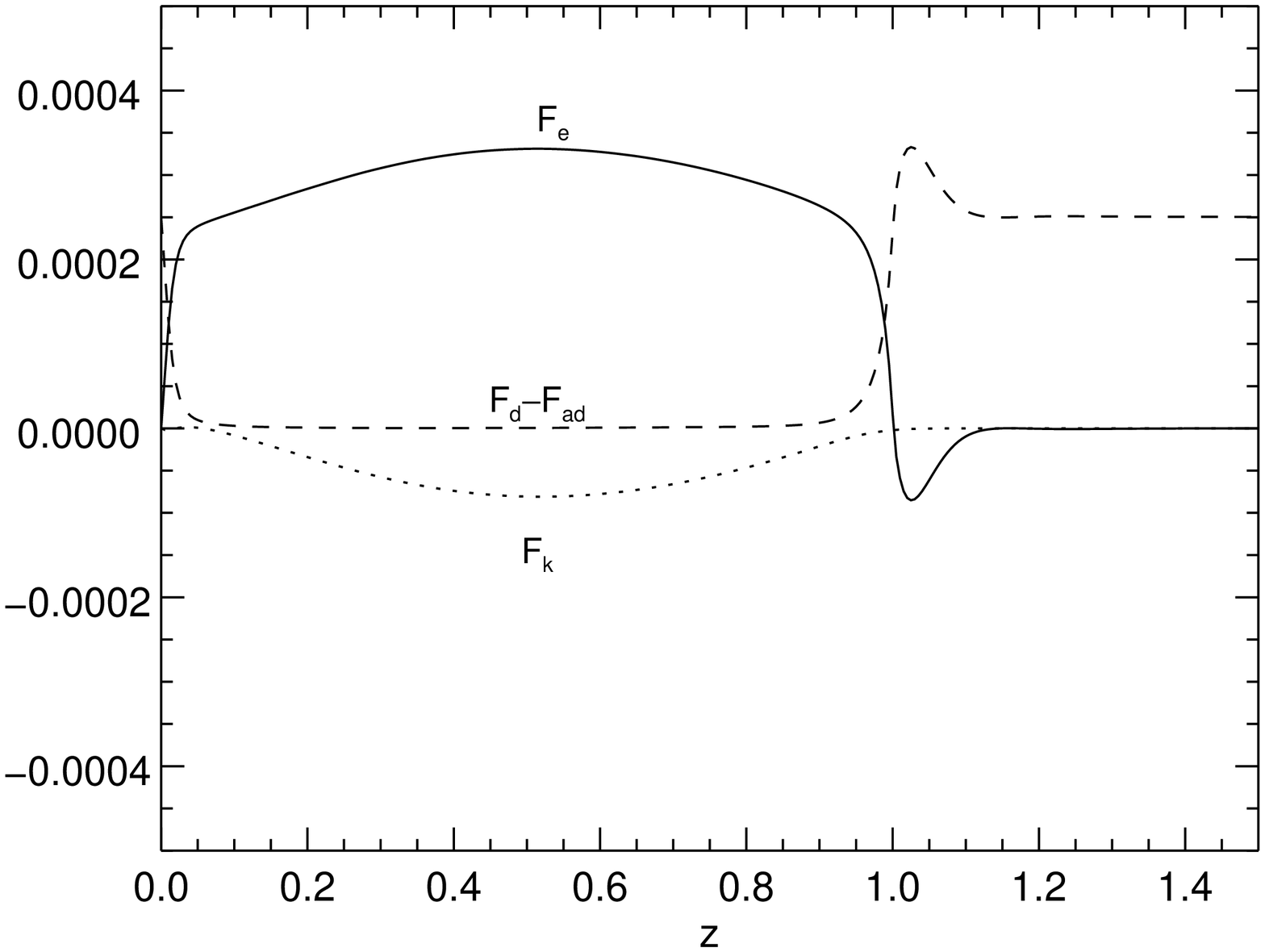}{0.5\textwidth}{(a)}
          \fig{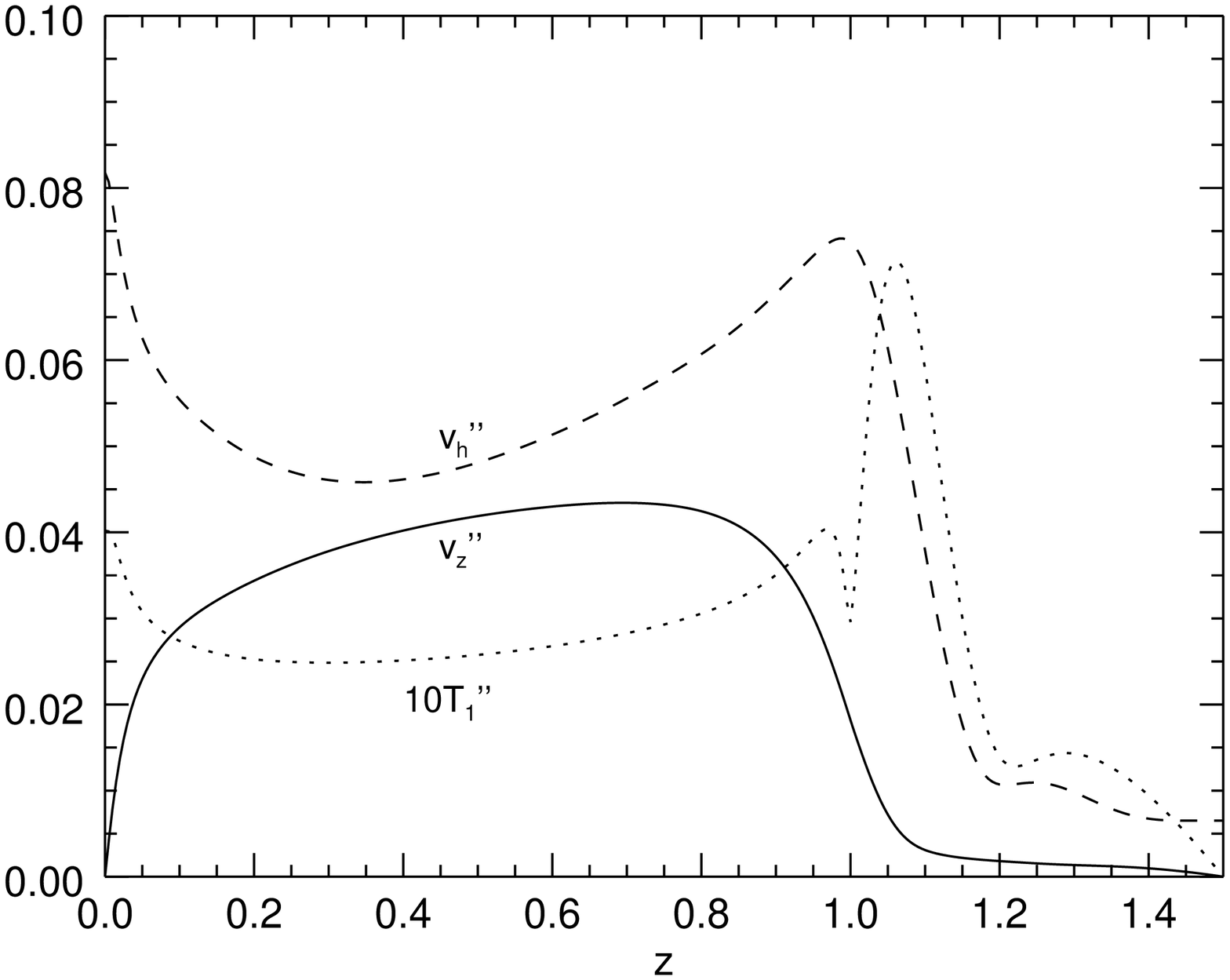}{0.5\textwidth}{(b)}
          }
\gridline{\fig{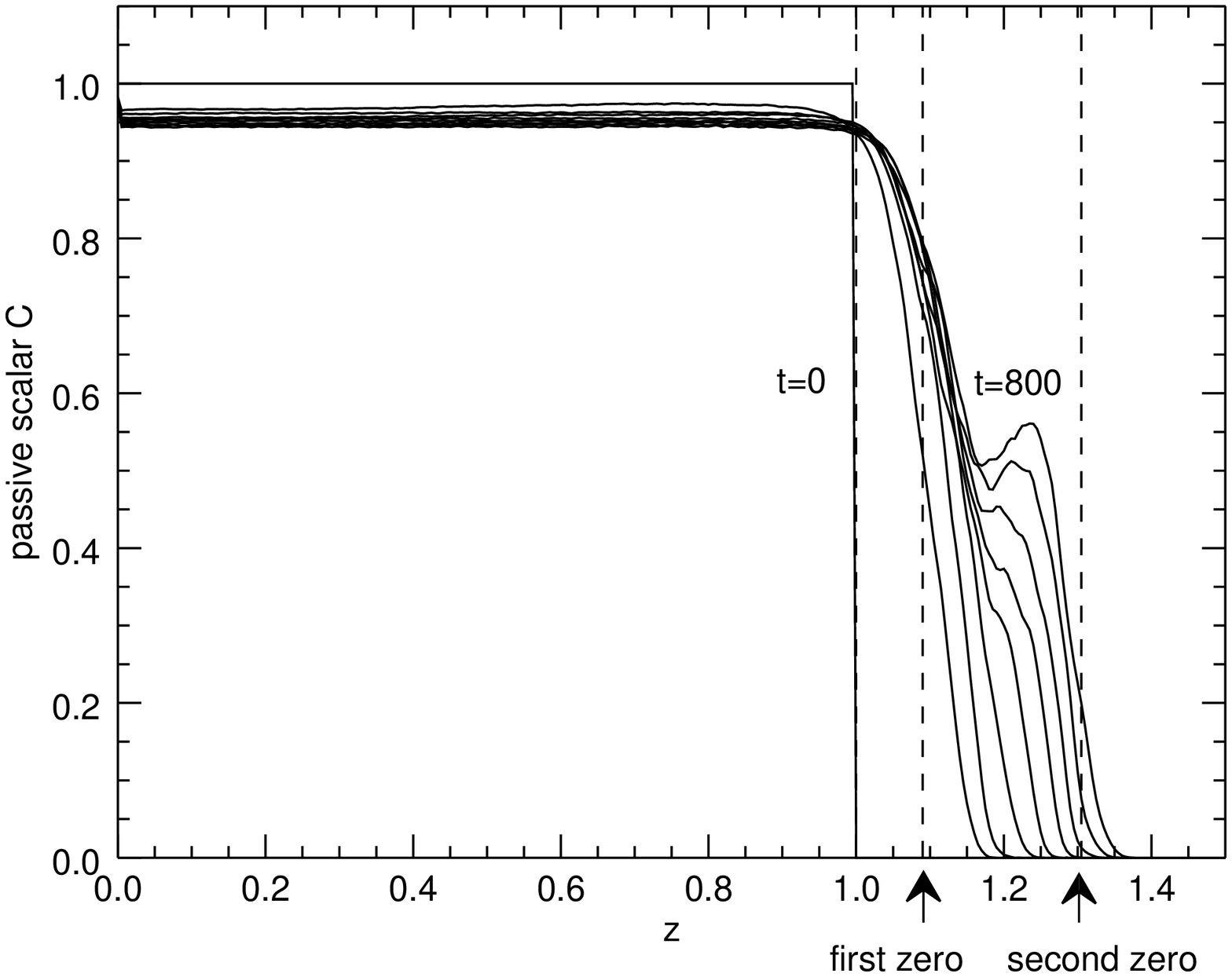}{0.5\textwidth}{(c)}
          \fig{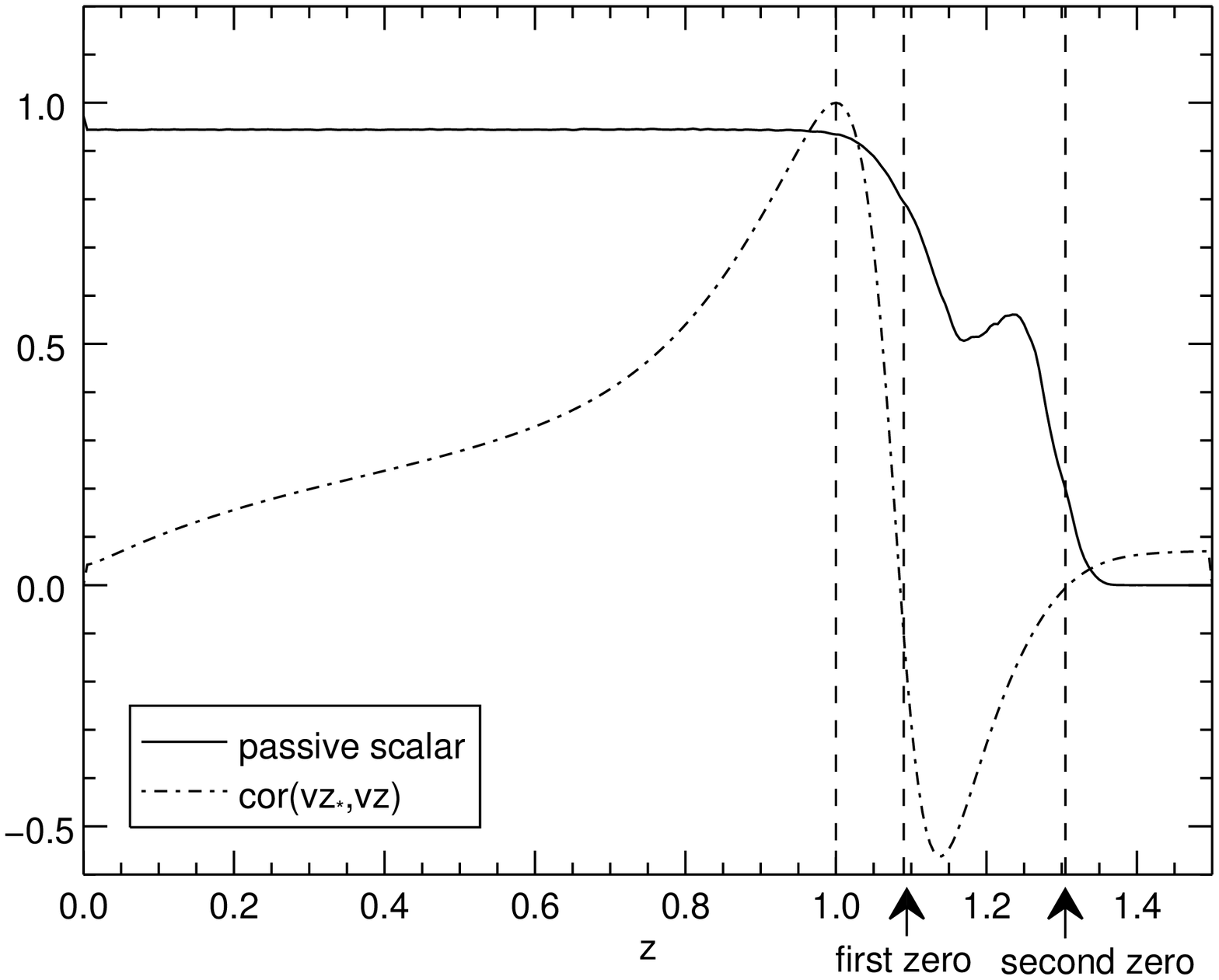}{0.5\textwidth}{(d)}
          }
\caption{The statistical results of the case A3. (a)The enthalpy flux $F_{e}$ (solid line), diffusive flux $F_{d}-F_{ad}$ (dashed line), and kinetic energy flux $F_{k}$ fluxes (dotted line) as functions of $z$. The adiabatic flux is deducted from diffusive flux. (b)The root mean square vertical velocity $v_{z}^{''}$ (solid line), horizontal velocity $v_{h}^{''}$ (dashed line),  and temperature perturbation $10T_{1}^{''}$ (dotted line) as functions of $z$. All the variables are taken averages both temporally and spatially (on the horizontal directions). (c) The evolution of the passive scalar concentration from t=0 to t=800. The passive scalar concentrations at t=0,100,200,300,400,500,600,700,800 (from left to right) are shown by solid lines. (d) The passive scalar concentration at t=800 is shown by the solid line. The correlation of vertical velocities between $v_{z}$ and $v_{z*}$ (vertical velocity at the interface) is shown by the dash-dotted line. The arrows indicate the locations of the first and second zero points of the veritical velocity correlations.\label{fig:f1}}
\end{figure}

The location of the overshooting distance below a convection zone (downward overshooting) is usually defined at the first zero point of the kinetic energy flux in the stable zone. However, it is not an appropriate proxy for the overshooting distance above a convection zone (upward overshooting). Unlike the downward overshooting, the penetrative flow is usually broad and weak in the upward overshooting region. In such case, the first zero point of the kinetic energy flux can hardly be identified, since it is too small and can be easily affected by numerical errors (see fig.~\ref{fig:f1}(a)). Thus we need find other indicators to measure the extent of upward overshooting distance. We plot the profiles of other possible candidates (the root mean square vertical velocity, horizontal velocity, temperature perturbation) on fig.~\ref{fig:f1}(b). From the figure, we see that the turning points of these physical variables are located at different places in the stable zone. The overshooting distance is not unique if these variables are used as indicators. A criteria to define the overshooting boundary is required. As the overshooting distance is highly related to the material mixing process in the stars, it is reasonable to assess the overshooting distance through measuring the transport of passive scalar. After the convection reaches a thermally relaxed state, we introduce passive tracer into the unstable zone, and then we evolve the system until the passive tracer penetrates into the stable zone and reaches a statistically steady state. The evolutionary equation of the passive scalar is
\begin{equation}
\frac{\partial \rho C}{\partial t}+\nabla\cdot (\mathbf{M}C)=0~,
\end{equation}
where the passive scalar $C$ takes an initial constant value of one in the unstable zone, and zero in the stable zone, respectively. After the system is relaxed, we compare the penetration of the passive scalar with the profiles of physical variables to find the best match. By this method, we have identified in previous simulations \citep{chan2010overshooting} that the second zero point of the velocity correlation (with the velocity at the interface) is a good indicator for the measurement of upward overshooting distance. {Here we perform a similar simulation on the case A3 for the illustrative purpose. Fig.\ref{fig:f1}(c) shows the evolutionary history of the passive scalar concentration from the time $t=0$ to the time $t=800$, where $t=0$ denotes the time when we introduced the passive scalar into the system. The passive scalar concentrations are shown in every 100 unit time by solid lines. Fig.\ref{fig:f1}(d) shows the passive scalar concentration at t=800, and the correlation coefficient of the vertical velocities $cor(v_{z*},v_{z})$ (here $v_{z*}$ means the vertical velocity at the interface). The passive scalar quickly penetrates into the convectively stable layer. From the fig.\ref{fig:f1}(c), we see that the mixing did not stop when the passive scalar reached the first zero point, although the correlation coefficient of vertical velocities turns to be negative above this location. The second zero point serves as a barrier for the mixing. The passive scalar concentration gradually accumulated after reaching the second zero point. Above the second zero point, mixing still happens but the process is much slower. The time scales of the turbulent dissipation and thermal dissipation are $\tau_{turb}=k/\varepsilon$ and $\tau_{ther}=(k^3/\varepsilon^2)(\rho c_{p}/\kappa_{i})$, respectively. Here $k$ is the kinetic energy, and $\varepsilon$ is the dissipation rate. Let $z_{1}$ and $z_{2}$ be the locations of the first zero point and the second zero point, respectively. The averaged values of $\tau_{turb}$ in the three regions $z\in(z_{0},z_{1})$, $z\in(z_{1},z_{2})$, $z\in(z_{2},z_{top})$ are 16.6, 8.2, 4.5 unit time. The averaged values of $\tau_{ther}$ in the corresponding regions are 1026.5, 8.5, 0.52. In the region $z\in(z_{0},z_{1})$, $\tau_{turb}$ is much smaller than $\tau_{ther}$. Thus the turbulent dissipation dominates the thermal dissipation. In the region $z\in(z_{1},z_{2})$, $\tau_{turb}$ is comparable to $\tau_{ther}$. Turbulent dissipation still plays a role in mixing material. Above $z_{2}$, $\tau_{turb}$ is several times larger than $\tau_{ther}$. Now the thermal dissipation dominates the turbulent dissipation. As a result, the material mixing is inactive above the location $z_{2}$. Thus we conclude that the second zero point of the velocity correlation is a good proxy for the measurement of the overshooting distance.}

\subsection{The effect of relative stability $S$}
\begin{figure}
\plotone{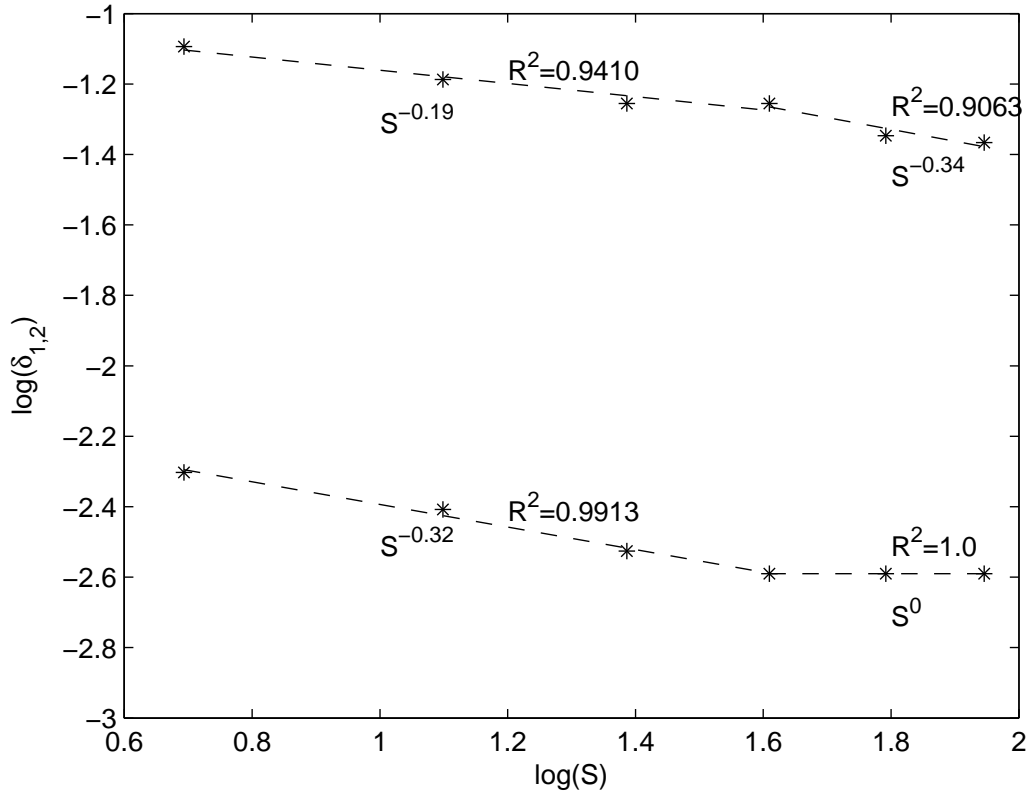}
\caption{Overshooting distance as a function of relative stability parameter $S$. The overshooting distances $\delta_{1}$ and $\delta_{2}$ of cases A2-A7 are shown with stars. The estimated scaling laws are presented with dashed lines. {The values of R-squared statistics $R^2$ are shown in the figure.} \label{fig:f2}}
\end{figure}

\begin{figure}
\plottwo{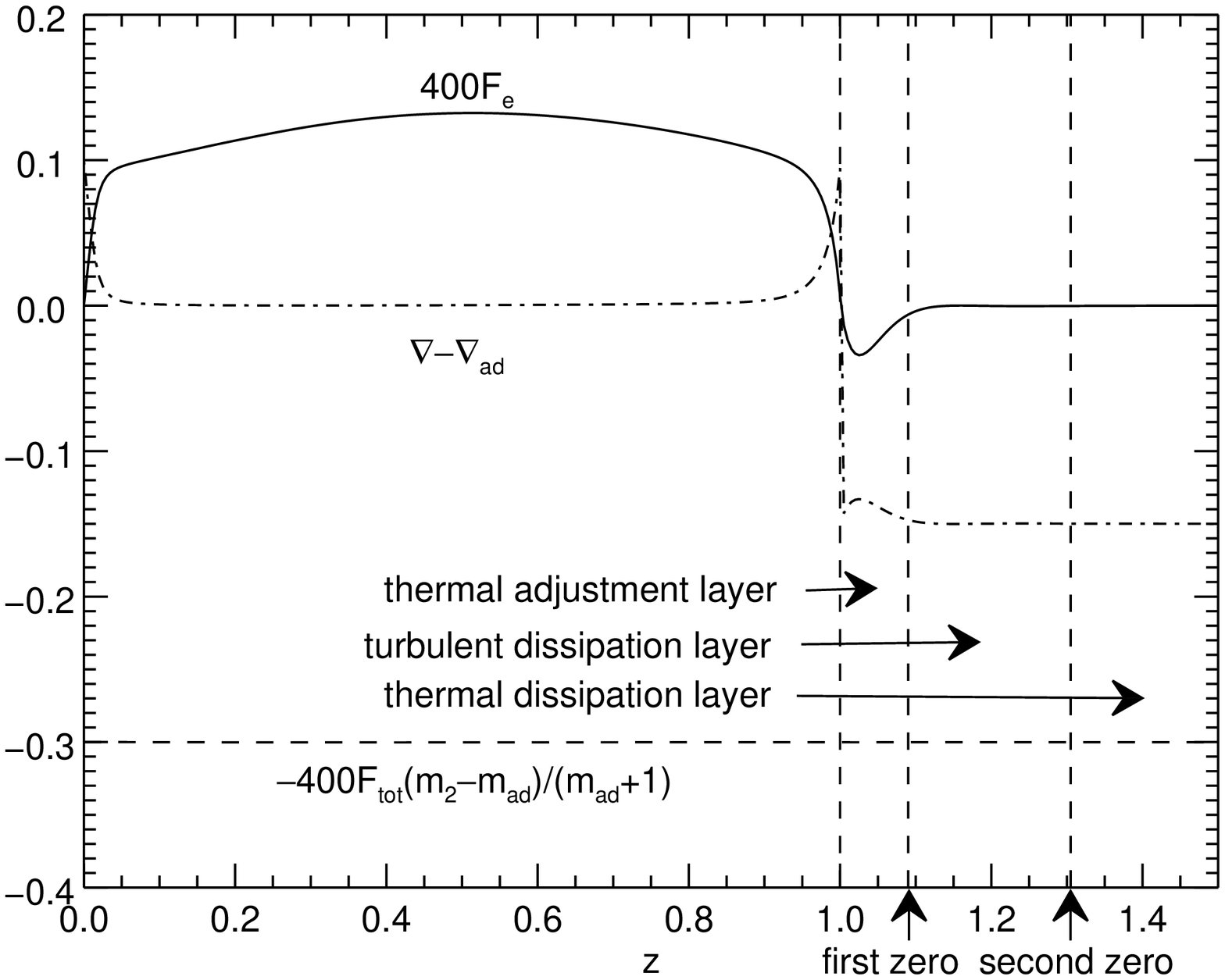}{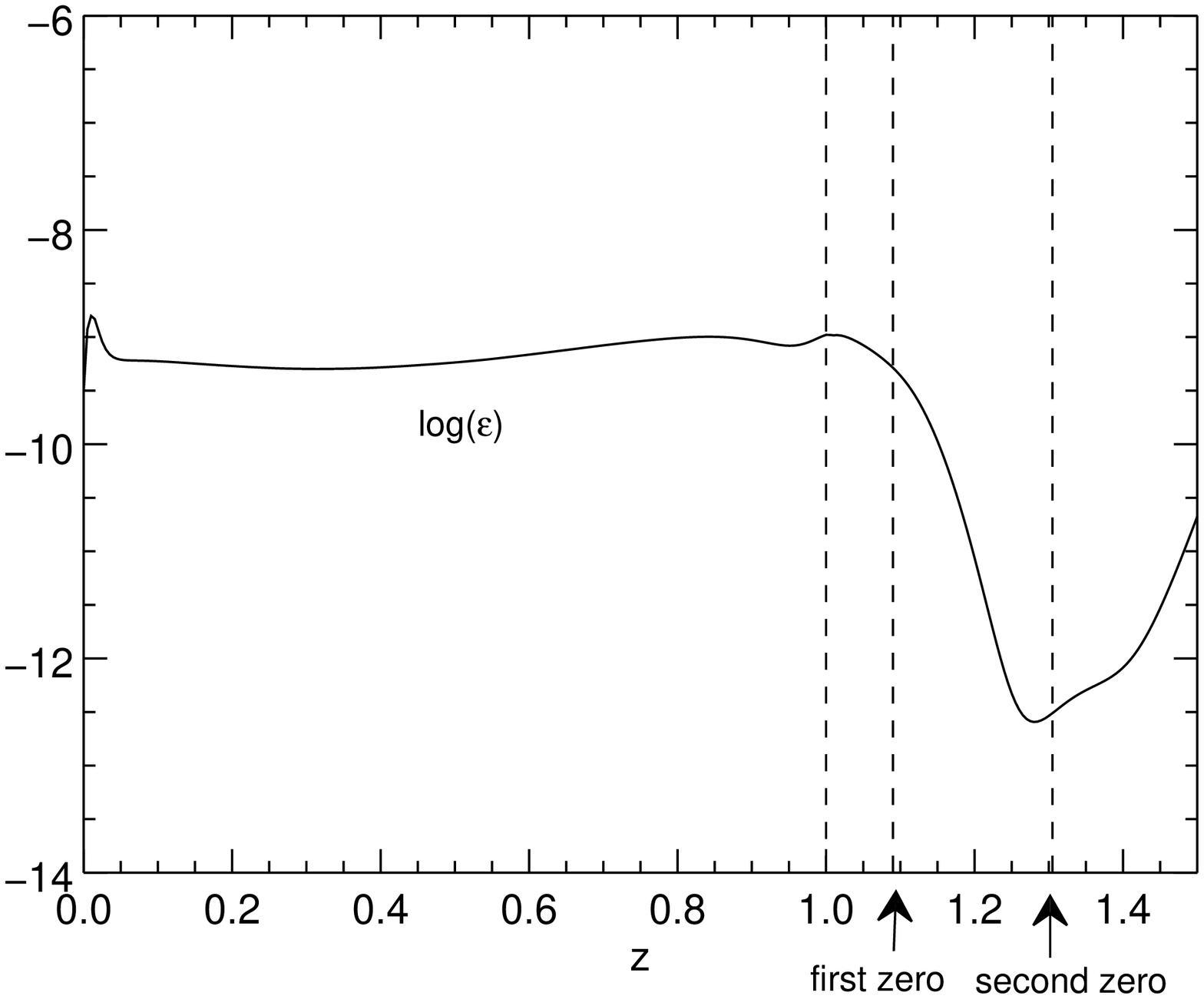}
\caption{The left panel shows the layer structure in the stable zone in the case A3. The right panel shows the logarithm of the kinetic energy dissipation rate $\varepsilon$ in the case A3. There are three layers above the convection zone: the thermal adjustment layer, the turbulent dissipation layer, and the thermal diffusion layer. The first zero point of velocity correlation locates the upper boundary of the thermal adjustment layer, and the second zero point locates the upper boundary of the turbulent dissipation layer.\label{fig:f3}}
\end{figure}

We now discuss the effect of relative stability parameter $S$ on overshooting distance. In this paper, the values $m_{1}=1$ and $m_{ad}=1.5$ are fixed. The polytropic index in the second layer $m_{2}=m_{ad}+(m_{ad}-m_{1})S$ increases with the value $S$. In this paper, we only focus on small values of $S$. Cases A1-A7 perform simulations with the relative stability parameter $S$ varying from 1 to 7, while keeping other parameters almost the same. The last two columns ($\delta_{1}$ and $\delta_{2}$) in table \ref{tab:table1} give the distances of the first and second zero points of velocity correlation, respectively. As mentioned above, we use the location of the second zero of velocity correlation to measure the overshooting distance, but here we also list the location of the first zero of velocity correlation for reference. To better illustrate the relationship, we show the fitted scaling laws of overshooting distances on parameter $S$ in the fig.~\ref{fig:f2}. Apparently, the scaling law of overshooting distance on $S$ is not unique. First, the scaling laws of $\delta_{1}$ and $\delta_{2}$ are different. Second, the scaling law of either $\delta_{1}$ or $\delta_{2}$ varies in different regimes of $S$. For the $\delta_{1}$, the scaling law is $\delta_{1}\sim S^{-0.32}$ in the smaller $S$ regime and $\delta_{1}\sim S^{0}$ in the higher $S$ regime. For the $\delta_{2}$, on the other hand, the scaling law is $\delta_{2}\sim S^{-0.19}$ in the smaller $S$ regime and $\delta_{2}\sim S^{-0.34}$ in the higher S regime. These scaling laws are different from the $S^{-1}$ and $S^{-0.25}$ scaling laws reported in the 2D \citep{hurlburt1994penetration} and 3D \citep{singh1995three,brummell2002penetration} numerical simulations of downward overshooting. \citet{hurlburt1994penetration} have explained the $S^{-1}$ and $S^{-0.25}$ scaling laws by the analytical model suggested by \citet{zahn1991convective}. According to Zahn's analysis, there exists two layers in the overshooting region: the nearly adiabatic penetration layer and the thermal adjustment layer. For the smaller $S$, the nearly adiabatic layer dominates and a scaling law of $S^{-1}$ is expected. For the higher $S$, on the other hand, the thermal adjustment layer dominates and a scaling law of $S^{-0.25}$ is expected. \citet{singh1995three} has confirmed the $S^{-1}$ and $S^{-0.25}$ laws in their large eddy simulations. \citet{brummell2002penetration}, however, has only identified the thermal adjustment layer and reported the $S^{-0.25}$ scaling law in their 3D direct numerical simulations. Zahn's analysis has been verified in the simulations of downward overshooting. However, the result of our upward overshooting simulations does not agree well with his analysis. The left panel of fig.~\ref{fig:f3} shows the layer structure in the upward stable zone. In the convection zone, the turbulent convection is very efficient and the temperature gradient is very close to the adiabatic temperature gradient. Above the convection zone, there is a thermal adjustment layer where the temperature gradient adjusts its value from adiabatic temperature gradient to radiative temperature gradient. In this layer, the buoyant force does a negative work and the convective flux turns to be negative. At the first zero point of velocity correlation, the temperature gradient almost finishes the thermal adjustment and the convective flux is nearly zero. The first zero point is approximately at the upper boundary of the thermal adjustment layer. Above the thermal adjustment layer, there is another turbulent dissipation layer. In this layer, although the mixing of entropy is negligible, the mixing of material is still active. The right panel of fig.~\ref{fig:f3} clearly shows that the dissipation rate of the turbulent kinetic energy $\varepsilon$ in this layer does not vanish until it reaches the second zero point. The turbulent dissipation layer is much wider than the thermal adjustment layer. The overshooting distance will be massively underestimated if the turbulent dissipation layer has not been considered. The interface at the second zero point serves as a barrier of material mixing. The passive scalar can hardly penetrate into the layer above this interface.

{In the left panel of fig.\ref{fig:f3}, we note that there is an abrupt change of the temperature gradient $\nabla$. In our initial settings, we set the temperature gradient to be piecewise constant. The abrupt change of $\nabla$ in the fig.\ref{fig:f3} could be a physical result or a numerical error. To justify it, we have run an additional simulation with more grid points near the interface. We find that the results have no big difference (see the appendix for more details). The abrupt change of $\nabla$ is also presented in the simulation results of the downward overshooting in \citet{korre2019convective} (see the figs.6,12,14 in their paper). Thus we believe that the abrupt change of $\nabla$ is a physical result instead of a numerical error.}

\subsubsection{The thermal adjustment layer}
\begin{figure}
\plottwo{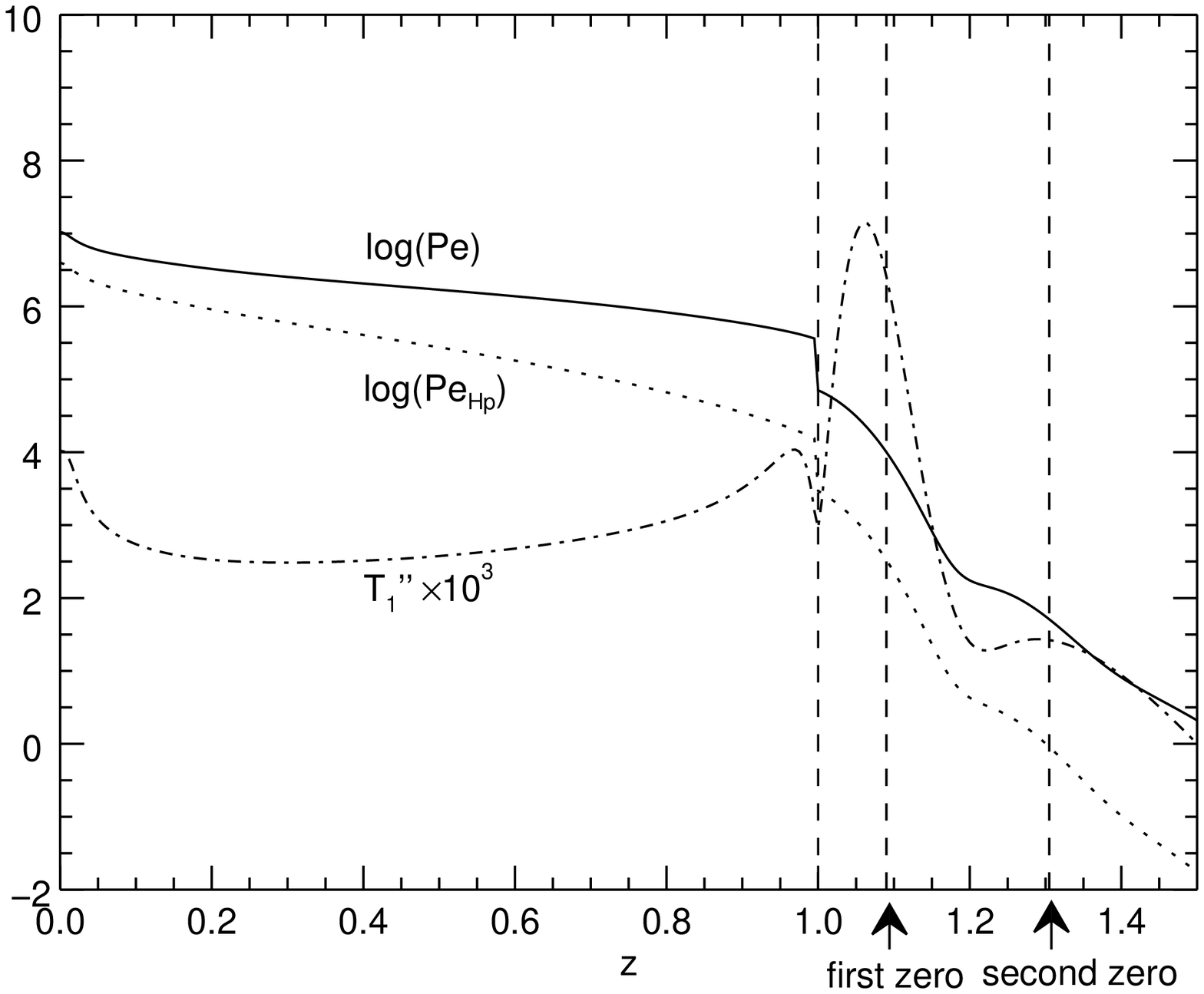}{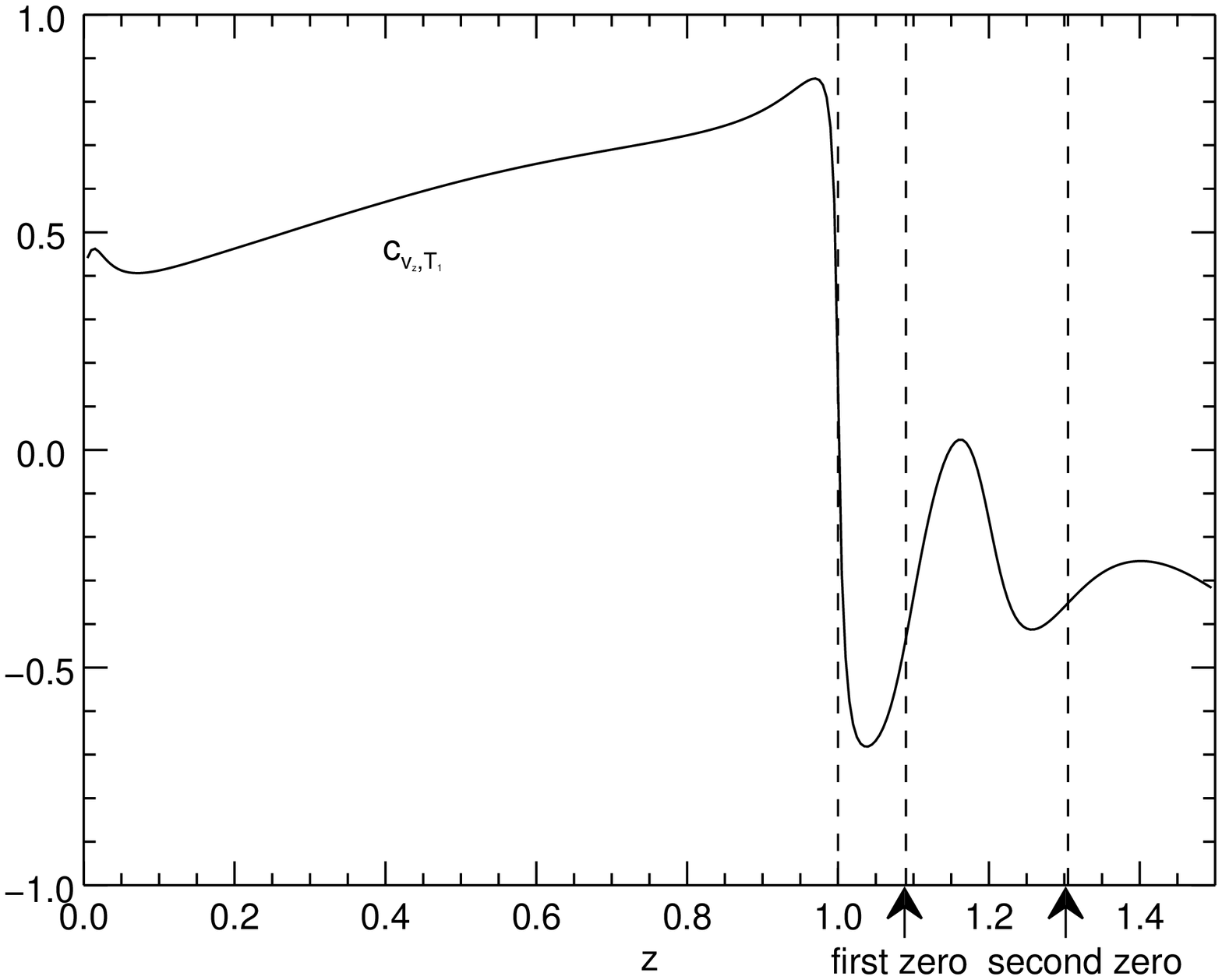}
\caption{Left: the profiles of P\'eclet number (solid curve for ${\rm Pe}$, and dotted curve for ${\rm Pe}_{Hp}$) and temperature perturbation (dash-dotted curve) in the case A3. ${\rm Pe}_{Hp}$ uses the local pressure height as the characteristic length. Right: the profile of the correlation coefficient $c_{v_{z},T_{1}}$ between the vertical velocity and the temperature perturbation in the case A3. \label{fig:f4}}
\end{figure}

Based on Zahn's analysis, \citet{hurlburt1994penetration} has estimated the thickness of the thermal adjustment layer in the downward overshooting zone. In this layer, three assumptions are made in their derivation. First, the P\'eclet number is of order unity. Second, the relative temperature perturbation decays from their initial value to zero. Third, the stratification is gradually changed from adiabatic to radiative, thus the enthalpy flux $F_{e}$ can be approximated with $-F_{tot}({m_{2}-m_{ad}})/({m_{ad}+1})$. However, these assumptions are not good at describing the thermal adjustment layer in the upward overshooting zone. First, the P\'eclet number is larger than one, and of order $100$ in this layer (the left panel of fig~\ref{fig:f4}). Second, the relative temperature perturbation first increases, and then decreases in this layer. {The value at the upper boundary of this layer is several times larger than that at the interface (the left panel of fig.~\ref{fig:f4}).} Third, the temperature gradient quickly adjusts from adiabatic to radiative within a very shallow region, thus the approximation $-F_{tot}({m_{2}-m_{ad}})/({m_{ad}+1})$ deviates a lot from the enthalpy flux $F_{e}$ (the left panel of fig.~\ref{fig:f3}). Therefore, the thickness estimated with these assumptions cannot be applied to upward overshooting.

To estimate the thickness of the thermal adjustment layer in the upward overshooting zone, we follow the fluid motions of the updraft. We make an assumption that the vertical velocity and the temperature perturbation are highly negatively correlated (the right panel of fig.\ref{fig:f4}). Following the work of \citet{zahn1991convective}, the deceleration of the updrafts can be described to the first order of density and temperature perturbations by
\begin{equation}
\frac{c}{2}\frac{dw^2}{dz}=-g\frac{\rho_{1}}{\rho}=g\frac{T_{1}}{T}~,\label{eq:draft}
\end{equation}
where $c$ is a coefficient measuring the degree of asymmetry of the flow. From fig.~\ref{fig:f3}, we find that the temperature gradient is close to radiative temperature gradient in this layer. Thus we assume that $\nabla-\nabla_{ad}=(\nabla-\nabla_{rad})+(\nabla_{rad}-\nabla_{ad})\approx (m_{ad}-m_{2})/[(m_{2}+1)(m_{ad}+1)]$ is constant in this region. Under these assumptions, we obtain the following differential equation
\begin{eqnarray}
\frac{c}{2}\frac{dw^2}{dz}&=&g\frac{T_{1}}{T}=-g\int_{1}^{z}\frac{1}{T}[\frac{\partial T}{\partial z'}-(\frac{\partial T}{\partial z'})_{ad}]dz' \nonumber\\
&=&  -g\int_{1}^{z}(\nabla-\nabla_{ad})\frac{d\log P}{dz'}dz' \approx g^2(\nabla-\nabla_{ad}) (z-1) \nonumber\\
&=&-g^2\frac{m_{ad}-m_{1}}{m_{ad}+1}\frac{S}{S(m_{ad}-m_{1})+(m_{ad}+1)}(z-1)~.
\end{eqnarray}
We further assume that the vertical velocity at the first zero point of velocity correlation is zero. By integrating the above equation from $z=1$ to $z=1+\delta_{1}$, we have
\begin{equation}
W_{*}^2\approx\frac{g^2}{c}\frac{m_{ad}-m_{1}}{m_{ad}+1}\frac{S}{S(m_{ad}-m_{1})+(m_{ad}+1)}\delta_{1}^2~,\label{eq18}
\end{equation}
where $W_{*}$ is the initial vertical velocity at the interface. Since the convection is efficient in the convection zone, we assume $F_{tot}-F_{ad}=\frac{m_{ad}-m_{1}}{m_{ad}+1}F_{tot} \sim \rho_{*} W_{*}^3$, where $\rho_{*}$ is the density at the interface. Under this assumption, we obtain the following equation
\begin{equation}
(\frac{F_{tot}}{\rho_{*}})^{1/3}\approx \frac{g}{c^{1/2}}(\frac{m_{ad}-m_{1}}{m_{ad}+1})^{1/6}[\frac{S}{S(m_{ad}-m_{1})+(m_{ad}+1)}]^{1/2}\delta_{1}~.\label{eq19}
\end{equation}
This equation implies two scaling laws. First, the distance $\delta_{1}$ and the total flux $F_{tot}$ obey a scaling law of $\delta_{1} \sim F_{tot}^{1/3}$. It has been confirmed in the large eddy simulations of overshooting above a convection zone \citep{chan2010overshooting}. Second, the distance $\delta_{1}$ and the relative stability parameter $S$ satisfy a scaling law of $\delta_{1}\sim \{{S}/[{S(m_{ad}-m_{1})+(m_{ad}+1)]}\}^{-1/2}$. Given $m_{ad}=1.5$ and $m_{1}=1$, the scaling law of S can be approximated by
\begin{equation}
\delta_{1} \sim (\frac{S}{0.5S+2.5})^{-1/2} \sim S^{-0.36}, \text{when $S\in\{1,2,3,4\}$}
\end{equation}
The fitted scaling index is close to that of the numerical simulations in the smaller $S$ regime ($S\in\{2,3,4\}$). The above estimation also indicates that $\delta_{1}$ is approximately a constant when $S$ approaches infinity. This trend has been verified in the simulations with $S\in\{5,6,7\}$. {The scalings are different from those derived in \citet{zahn1991convective} and \citet{hurlburt1994penetration}, although the depicted physical pictures are the same (see eq.(\ref{eq:draft})). The difference comes from the different assumptions on the thermal structure in this layer. \citet{zahn1991convective} and \citet{hurlburt1994penetration} made the assumption of $\nabla\approx \nabla_{ad}$, whereas we assume $\nabla\approx \nabla_{rad}$ here in this work. As a result, the estimated values of the temperature perturbations on the right hand side of eq.(\ref{eq:draft}) are different.}

\subsubsection{The turbulent dissipation layer}
\begin{figure}
\plotone{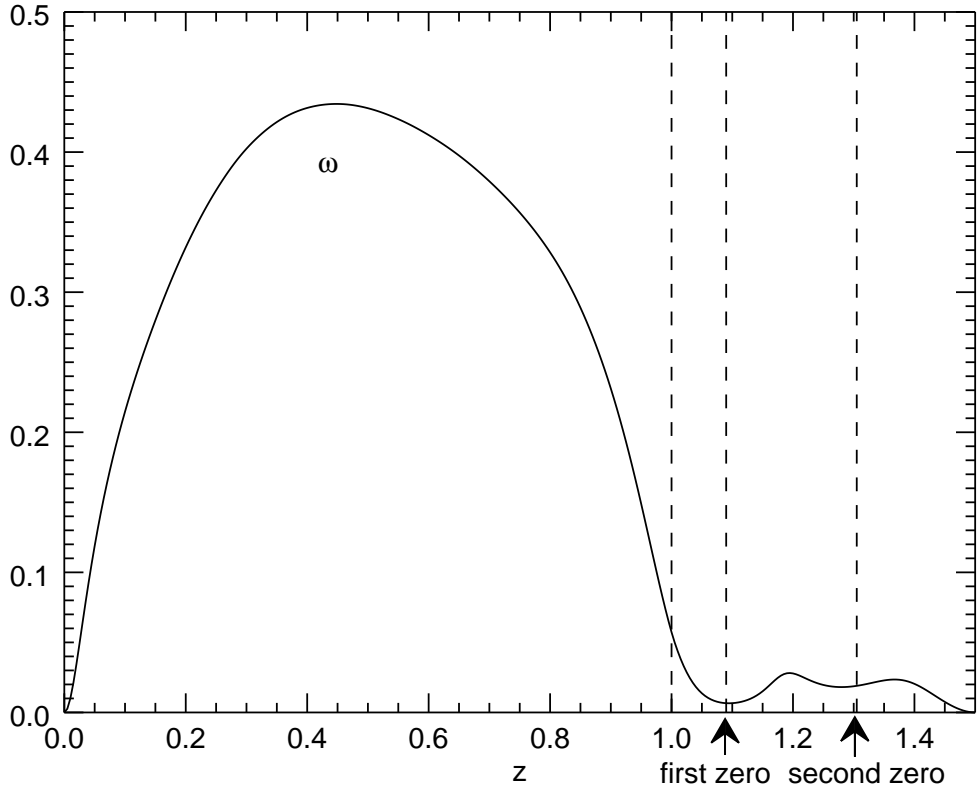}
\caption{The profile of the anisotropic degree $\omega$ in the case A3. \label{fig:f5}}
\end{figure}

From a turbulent convection model, \citet{zhang2012turbulent} discovered that a turbulent dissipation layer exists above the turbulent heat flux overshooting region. They made three assumptions in their derivation. First, the P\'eclet number ${\rm Pe}>>1$. Second, all variations except the turbulent perturbations are constant. Third, $\nabla\approx\nabla_{rad}$ at the place far away from the convective boundary. With these assumptions, they concluded that the boundary of the turbulent heat flux overshooting region is at the peak of $T_{1}''$, and the boundary of the turbulent dissipation layer is at the location of ${\rm Pe}=1$ \citep{zhang2012turbulent}. The anisotropic degree $\omega$, defined as the ratio of the vertical kinetic energy to the total kinetic energy, plays an important role in this turbulent dissipation layer \citep{zhang2012turbulent}. In their calculation, they found that $\omega$ almost equals to an constant equilibrium value in the convection zone with $\omega_{cz}>1/3$, because the buoyancy boosts the vertical fluid motions. They also showed that $\omega$ achieves an equilibrium value in the overshooting zone with $\omega_{ov}<1/3$, as now the buoyancy prevents the vertical fluid motions \citep{zhang2012turbulent}. Their analysis agrees well with our numerical result in several aspects. First, the assumptions they made are reasonable, which are verified in fig.~\ref{fig:f3} and fig.~\ref{fig:f4}. Second, the upper boundary of the thermal adjustment layer (the first zero point) in our simulation is close to the peak of $T_{1}''$ (fig~.\ref{fig:f4}). Third, the anisotropic degree indeed almost equals to an equilibrium value at the overshooting zone (fig~.\ref{fig:f5}). Despite these remarkable agreements, the boundary location of the turbulent dissipation layer seems questionable. From the left panel of fig.~\ref{fig:f4}, we see that the second zero point is not located at the ${\rm Pe}=1$ as predicted by \citet{zhang2012turbulent}. The reason is that the definitions of ${\rm Pe}$ are different. The local pressure scale height $H_{p}$ instead of $L_{1z}$ is chosen as the characteristic length for ${\rm Pe}$ in \citet{zhang2012turbulent}. Replacing $L_{1z}$ with $H_{p}$, fig.~\ref{fig:f4} clearly shows that the location of ${\rm Pe}_{Hp}=1$ agrees well with the second zero point. To estimate the thickness of the turbulent dissipation layer, we write
\begin{equation}
{\rm Pe}_{Hp}=\frac{\rho c_{p} v'' H_{p}}{\kappa_{2}}= \frac{c_{p} p v''}{F_{tot}(m_{2}+1)}\sim 1~.\label{eq21}
\end{equation}
The asymptotic solution of the turbulent convection model in the overshooting region \citep{zhang2012turbulent} gives
\begin{equation}
v''\sim c_{0} p^{\beta}
\end{equation}
where $c_{0}$ and $\beta$ are constants. The fitted values in the case A3 are $c_{0}=0.05$ and $\beta=1.16$. Substituting $v''$ into ${\rm Pe}_{Hp}$, we obtain
\begin{equation}
\delta_{2}=\frac{m_{2}+1}{g}[1-(\frac{F_{tot}(m_{2}+1)}{c_{p}c_{0}})^{(1/(\beta+1)(m_{2}+1))}]~.
\end{equation}
Given that $m_{2}=0.5S+1.5$, we find the approximated scaling laws are
\begin{equation}
\delta_{2} \sim S^{-0.19}, \text{when $S\in\{2,3,4,5\}$}~,
\end{equation}
and
\begin{equation}
\delta_{2} \sim S^{-0.34}, \text{when $S\in\{5,6,7\}$}~,
\end{equation}
respectively. This result agrees well with the scaling laws estimated from the second zero points.

\subsection{The effect of Prandtl number ${\rm Pr}$}
\begin{figure*}
\gridline{\fig{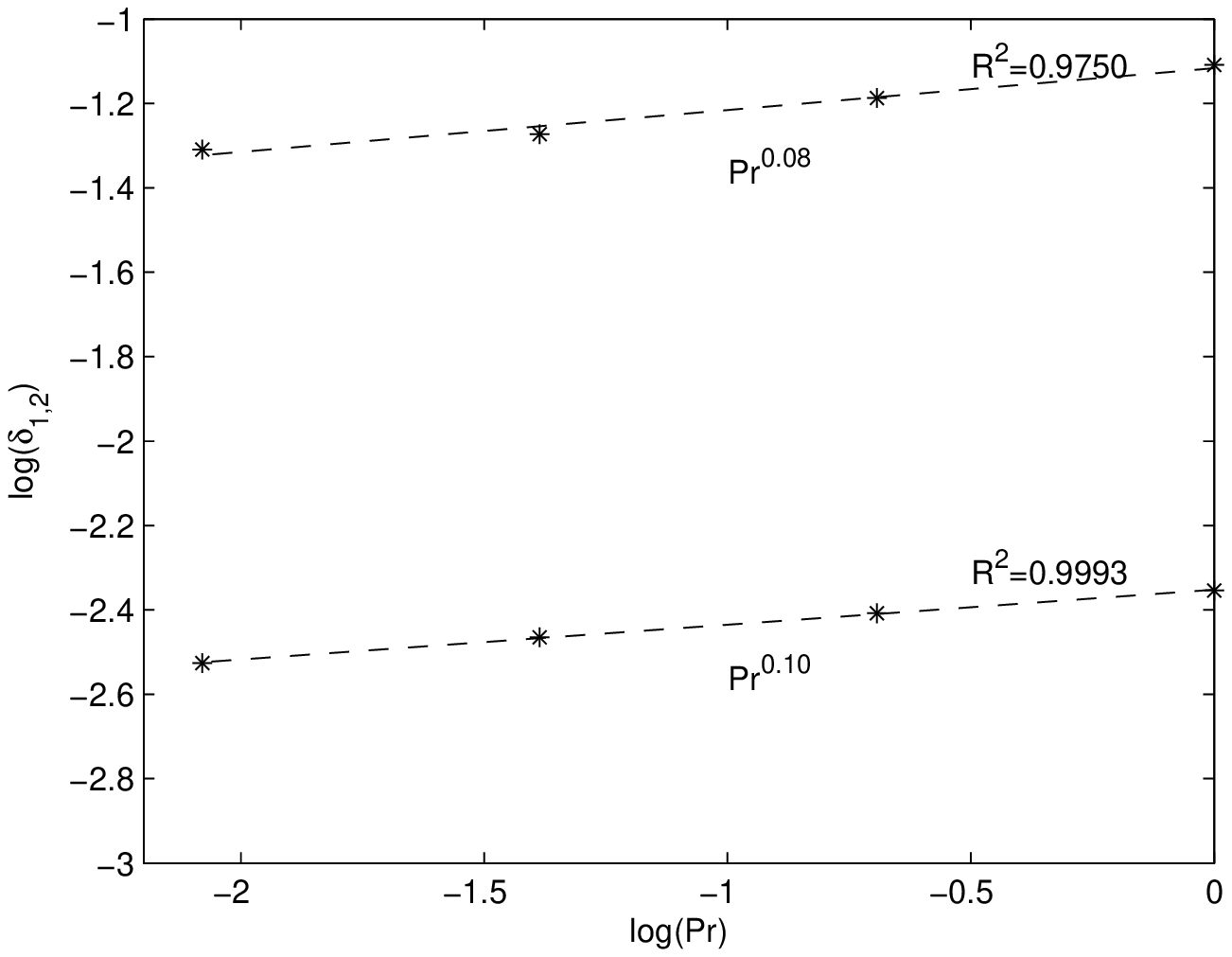}{0.5\textwidth}{(a)}
          \fig{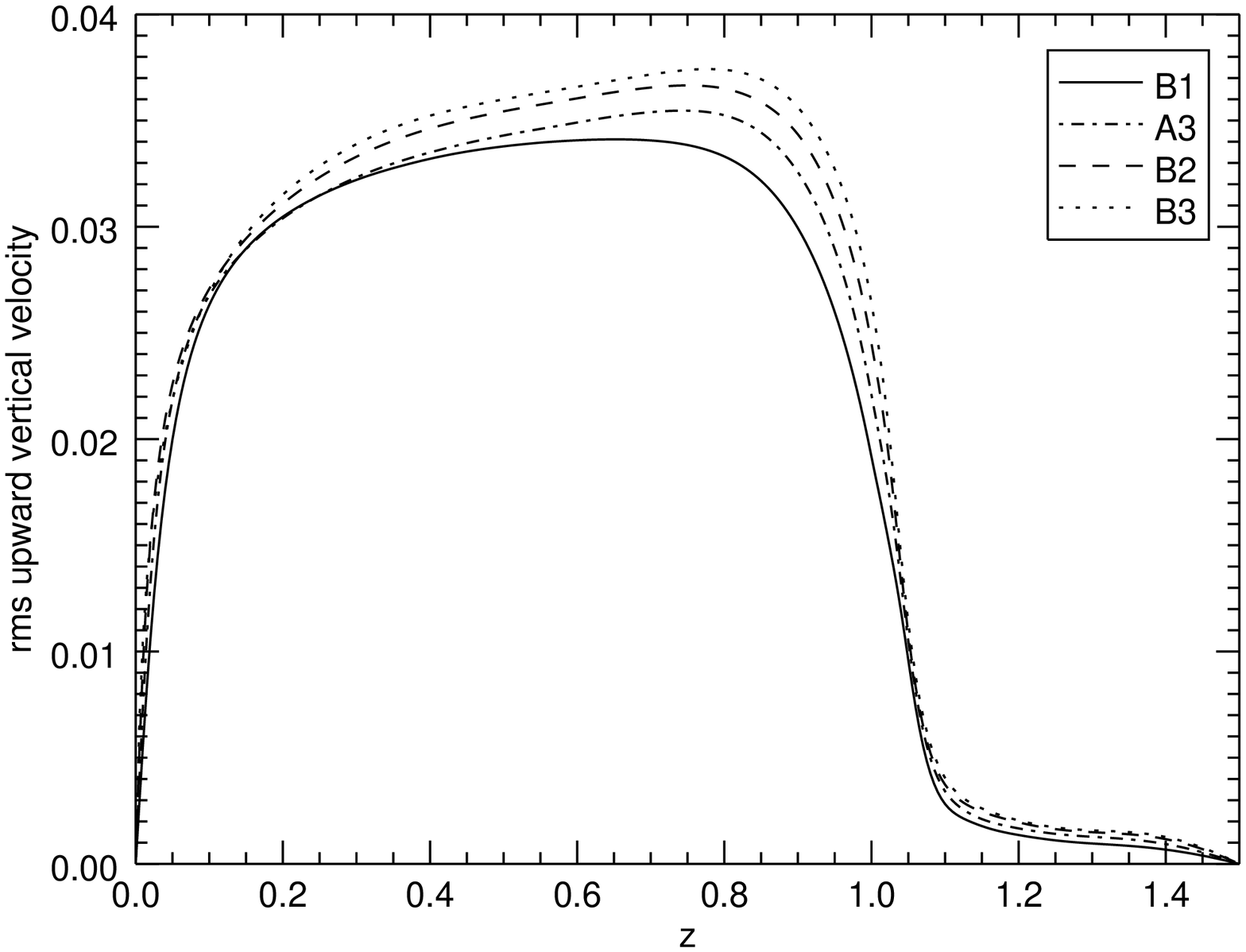}{0.5\textwidth}{(b)}
          }
\gridline{\fig{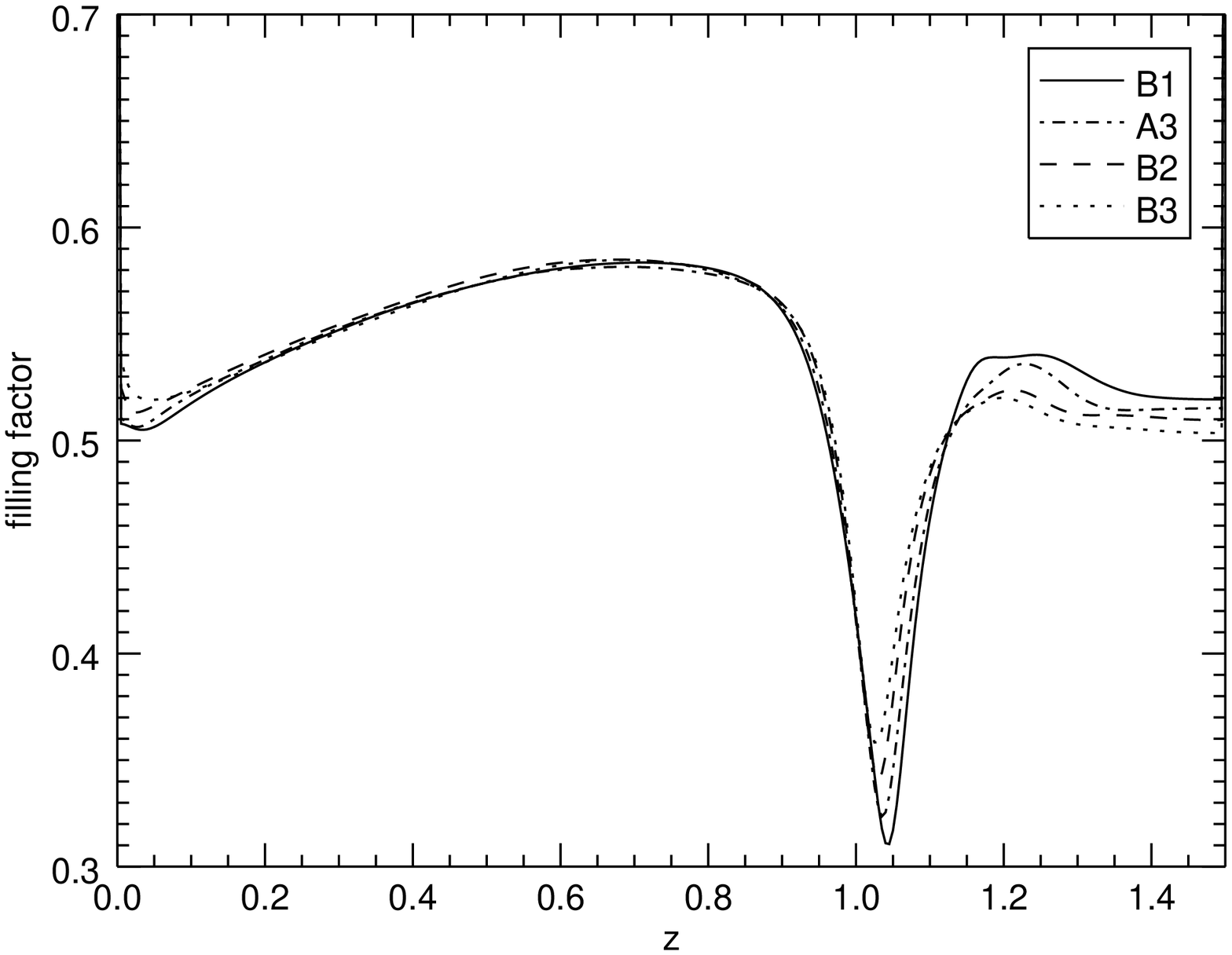}{0.5\textwidth}{(c)}
          \fig{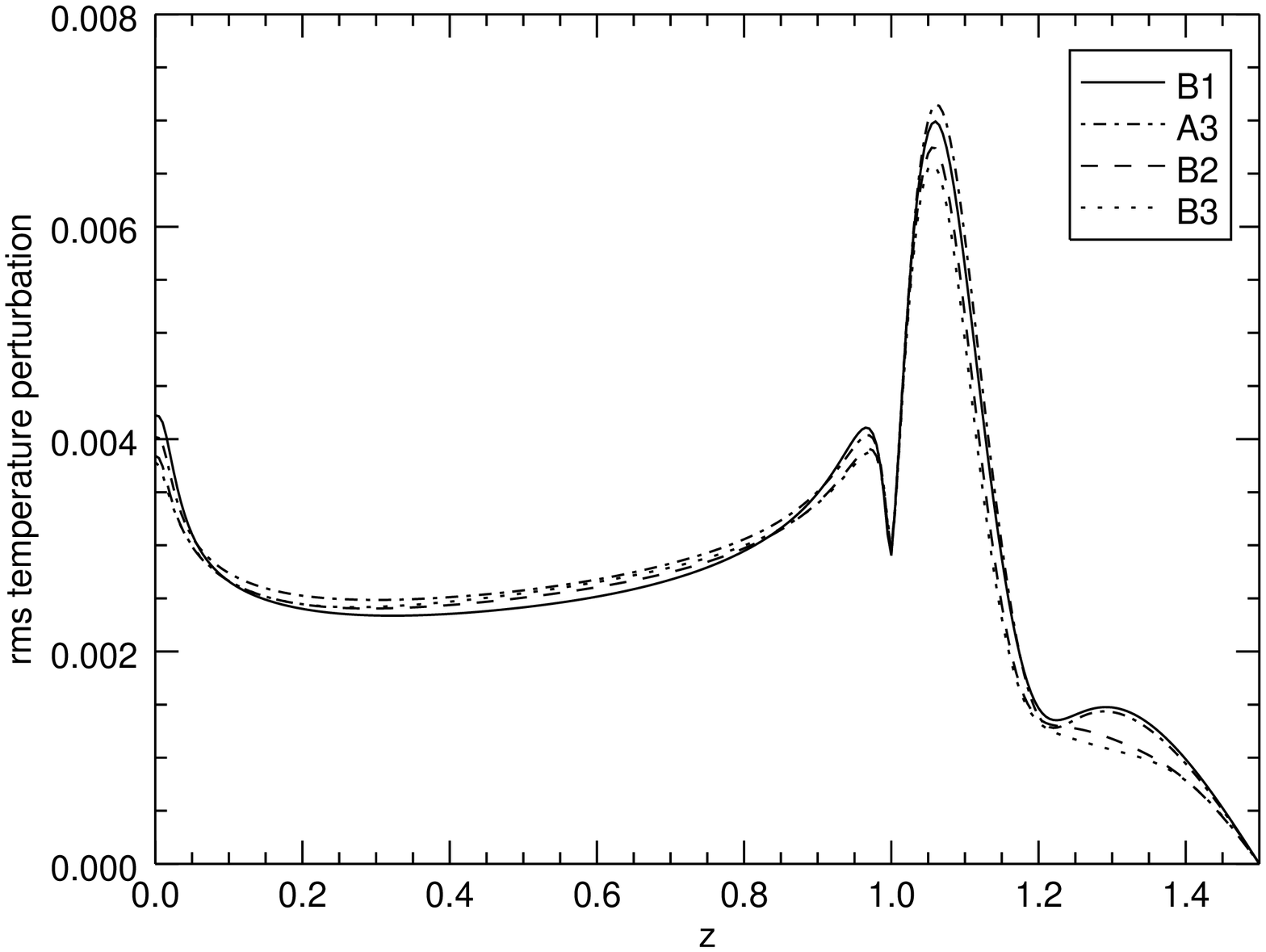}{0.5\textwidth}{(d)}
          }
\caption{(a)Overshooting distance as a function of Prandtl number ${\rm Pr}$. The overshooting distances $\delta_{1}$ and $\delta_{2}$ of cases A3, B1-B3 are shown with stars. The estimated scaling laws are presented with dashed lines. (b)Profile of root mean square upward vertical velocity. (c)Filling factor of upward flows. (d)Profile of root mean square temperature perturbation.  \label{fig:f6}}
\end{figure*}

In this subsection, we consider the effect of $\rm Pr$ on overshooting distance by comparing cases B1, A3, B2, and B3. The dependence of overshooting distances on $\rm Pr$ is shown on fig.~\ref{fig:f6}(a). The depth of thermal adjustment layer has a scaling law of $\delta_{1}\sim {\rm Pr}^{0.1}$, and that of turbulent dissipation layer has a scaling law of $\delta_{2}\sim {\rm Pr}^{0.08}$. Somewhat counter-intuitively, both scaling laws indicate that the overshooting distance increases with $\rm Pr$ (or viscosity). Lower viscosity leads to higher Reynolds number and root mean square velocity in the convection zone (see table~\ref{tab:table1}), thus one might expect that the overshooting distance decreases with $\rm Pr$. In the previous section, it has been shown that the overshooting distance highly depends on the upward vertical velocity at the interface between unstable and stable zones. Probably the upward vertical velocity at the interface is stronger at the lower $\rm Pr$ cases. However, fig.~\ref{fig:f6}(b) reveals that the upward vertical velocity decreases with $\rm Pr$. In the simulations of overshooting below a convection zone, \citet{brummell2002penetration} discovered a similar relationship between overshooting distance and Prandtl number $\delta\sim\rm Pr^{0.05}$. They attributed the positive correlation to the filling factor. Unfortunately, that does not seem to be the case in the upward overshooting. Fig.\ref{fig:f6}(c) clearly shows that the filling factor decreases with $\rm Pr$ in the thermal adjustment layer. The higher $\rm Pr$, the fewer upward flows penetrate into the overshooting zone through the interface. Thus the filling factor cannot explain the increase of the overshooting distance in the higher $\rm Pr$ cases. From a Reynolds stress model, \citet{xiong2001structure} has explained that the nonlocal effect of turbulent motions in the overshooting zone is also important. Let us consider a hot convective eddy penetrating upward through the interface. When it passes through the interface, it will keep moving upward and cool down. As the eddy moves upward, the temperature perturbation changes sign to be negative and its magnitude increases (see fig.\ref{fig:f1}). Keep moving upward, the eddy will blend into the surroundings when it reaches the boundary of the thermal adjustment layer, where the temperature of the eddy is the same with that of environment. From this simple description, we understand that there is a peak of temperature perturbation in the overshooting zone (see fig.\ref{fig:f6}(d)). Since the magnitude of the temperature perturbation changes abruptly in this region, the nonlocal turbulent diffusive effect is also important. The higher viscosity, the stronger nonlocal turbulent diffusive effect. Therefore the magnitude of temperature perturbation increases with $\rm Pr$ (see fig.\ref{fig:f6}(d)). Previous discussion shows that the thickness of overshooting is closely related to the peak of temperature perturbation. As a result, the thickness of the thermal adjustment layer increases with $\rm Pr$. The thickness of the turbulent dissipation layer, on the other hand, probably depends on the filling factor. As seen from fig.\ref{fig:f6}(c), the filling factor increases with $\rm Pr$ in this layer. The overshooting distance increases when the filling factor increases.

\subsection{The effect of P\'eclet number ${\rm Pe}$}
\begin{figure}
\plotone{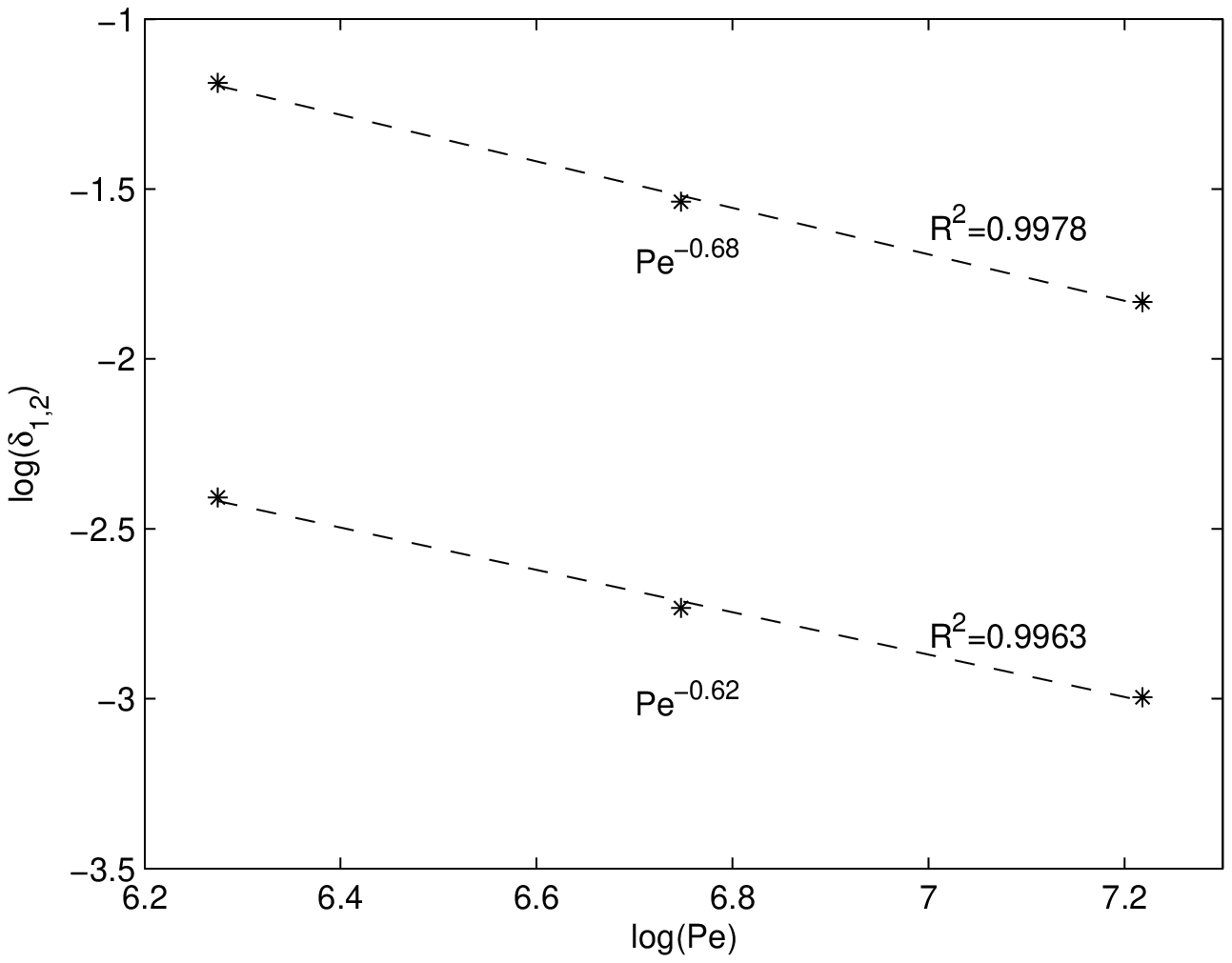}
\caption{Overshooting distance as a function of P\'eclet number ${\rm Pe}$. The overshooting distances $\delta_{1}$ and $\delta_{2}$ of cases A3, C2-C3 are shown with stars. The estimated scaling laws are presented with dashed lines. \label{fig:f7}}
\end{figure}

Now we discuss the effect of $\rm Pe$ on the overshooting distance by comparing cases C1, A3, C2, and C3. In this group, the total flux and viscosity are varied to make $\rm Pr$ to be constant. From eq.(\ref{eq18}) and eq.(\ref{eq19}), we can estimate that $W_{*}$ and $F_{tot}$ have the scalings of $W_{*}\sim \delta_{1}$ and $F_{tot}\sim \delta_{1}^{3}$, respectively. From the definition of $\rm Pe$, we get $\rm Pe$ in convection zone:
\begin{equation}
{\rm Pe}=\frac{\rho c_{p}\langle v'' \rangle_{cz} L_{1z}}{\kappa_{1}}=\frac{\rho c_{p} \langle v'' \rangle_{cz} L_{1z}g}{F_{tot}(m_{1}+1)}\sim \delta_{1}^{-2}
\end{equation}
where $\langle v'' \rangle_{cz}$ is the averaged value in the convection zone. Therefore we have a scaling law of $\delta_{1}\sim {\rm Pe}^{-0.5}$. However, this scaling law has not taken into account the effect from viscosity. The effect of viscosity on $\delta_{1}$ is approximately $\delta_{1} \sim \mu^{0.1}$ (see the previous subsection). Taking this into account, we obtain a scaling law of $\delta_{1}\sim {\rm Pe}^{-0.6}$. Fig.\ref{fig:f7} shows the fitted scaling law of overshooting distance on $\rm Pe$. The two scalings of $\delta_{1}$ and $\delta_{2}$ are $\delta_{1}\sim {\rm Pe^{-0.62}}$ and $\delta_{2}\sim {\rm Pe^{-0.68}}$, respectively. The scaling law of $\delta_{1}$ is very close to that of the analysis. It is not easy to deduce the scaling law of $\delta_{2}$ from eq.~(\ref{eq21}), since $c_{0}$ and $\beta$ vary too much when the total flux is changed. Based on the simulation result, it seems that the scalings of $\delta_{2}$ and $\delta_{1}$ are quite similar.

\section{Summary}
In this paper, we present simulation results of upward overshooting in turbulent compressible convection. The effects of stability parameter, the Prandtl number, and the P\'eclet number are considered. One major question on stellar structure and evolution is on how to determine the location of the overshooting boundary. In our previous work \citep{chan2010overshooting}, we have shown that the second zero point of the velocity correlation (with the velocity at the interface) is a good proxy to locate the position of the overshooting boundary. This study extends our knowledge on the physical relations between the zero points and overshooting mixing. The present work reveals that the convectively stable zone above a convection zone can be partitioned into three layers: the thermal adjustment layer (mixing both entropy and material), the turbulent dissipation layer (mixing material but not entropy), and the thermal dissipation layer (mixing neither entropy nor material). The first and second zero points of velocity correlation separate the turbulent dissipation layer from the thermal adjustment layer and the thermal dissipation layer. The overshooting zone should include both the thermal adjustment layer and the turbulent dissipation layer. This layer structure is significantly different from that of overshooting below a convection zone. For example, only a thermal adjustment layer is discovered in the 3D simulations of downward overshooting \citep{brummell2002penetration}. It is also different from the layer structure described by \citet{zahn1991convective} in his analytical nonlocal model. The simulation shows that the temperature perturbation tends to increase and $\rm Pe>1$ in the thermal adjustment layer, which are against the assumptions made in \citet{zahn1991convective}. However, our result shows agreement in many aspects with the analysis carried out by \citet{zhang2012turbulent} in their RSM. First, the layer structure is similar to their prediction. Second, the upper boundary of the adjustment layer is close to the peak of the magnitude of the temperature perturbation. Third, the P\'eclet number at the upper boundary of the turbulent dissipation layer is close to 1. Despite of the complexity on numerical calculation, the RSM has the advantage on the prediction of overshooting distance.

We have also examined the scaling laws of overshooting distance on the physical parameters $S$, $\rm Pr$, and $\rm Pe$. The power index fitted on $S$ is not unique. The trend toward higher $S$ indicates that the thickness of the thermal adjustment layer tends to be constant. However, the thickness of turbulent dissipation layer tends to decreases faster in the higher $S$ regime. As only several data points are collected, this trend could be seriously affected by fitting errors. Conclusive conclusion requires more simulation cases in the high $S$ regime. We plan to discuss it in our future work. The scaling law on $\rm Pr$ shows that the overshooting distance increases with the Prandtl number. Although similar trend has been identified in the downward overshooting, the mechanism is different. In the downward overshooting, this positive correlation is due to the increasing filling factor \citep{brummell2002penetration}. However, our calculation shows that the nonlocal effect of turbulent diffusion, instead of the filling factor, is the key factor on facilitating upward overshooting. Finally, we have discussed the scaling law of overshooting distance on $\rm Pe$. The result shows a decreasing trend of overshooting distance on increasing $\rm Pe$.

In terms of the local pressure scale height at the interface ($H_{p*}$), $\delta_{1}$ and $\delta_{2}$ of our simulations are in ranges of $[0.2,0.4]H_{p*}$ and $[0.64,1.34]H_{p*}$, respectively. The cases A1 and C1 are excluded in the discussion since the computational domain is not high enough. The ratio of $(\delta_{2}-\delta_{1})/\delta_{1}$ measures the relative thickness of the turbulent dissipation layer to the thermal adjustment layer. For all of our simulations, this ratio is almost a constant with a mean value of 2.42 and a standard deviation of 0.15. As mentioned above, the turbulent dissipation layer plays an important role in mixing material. It will massively underestimate the extent of overshooting distance if this layer is excluded. A reasonable estimation of the overshooting distance is crucial for the evolution and structure of stars with convective cores. This simple ratio might be useful in modelling core overshooting in stellar evolutionary code.

In the current paper, a step function of the heat conductivity is used in our initial settings. In the real stars, the heat conductivity changes gradually. Recent studies \citep{kapyla2018overshooting,korre2019convective} on the downward overshooting have shown that the scaling exponent of overshooting distance depends on the smoothness of the transition of the heat conduction profile. It is worthwhile to consider the effect of the smoothness of the transition in the upward overshooting. We will investigate this effect in the future work.

\begin{figure}
\plotone{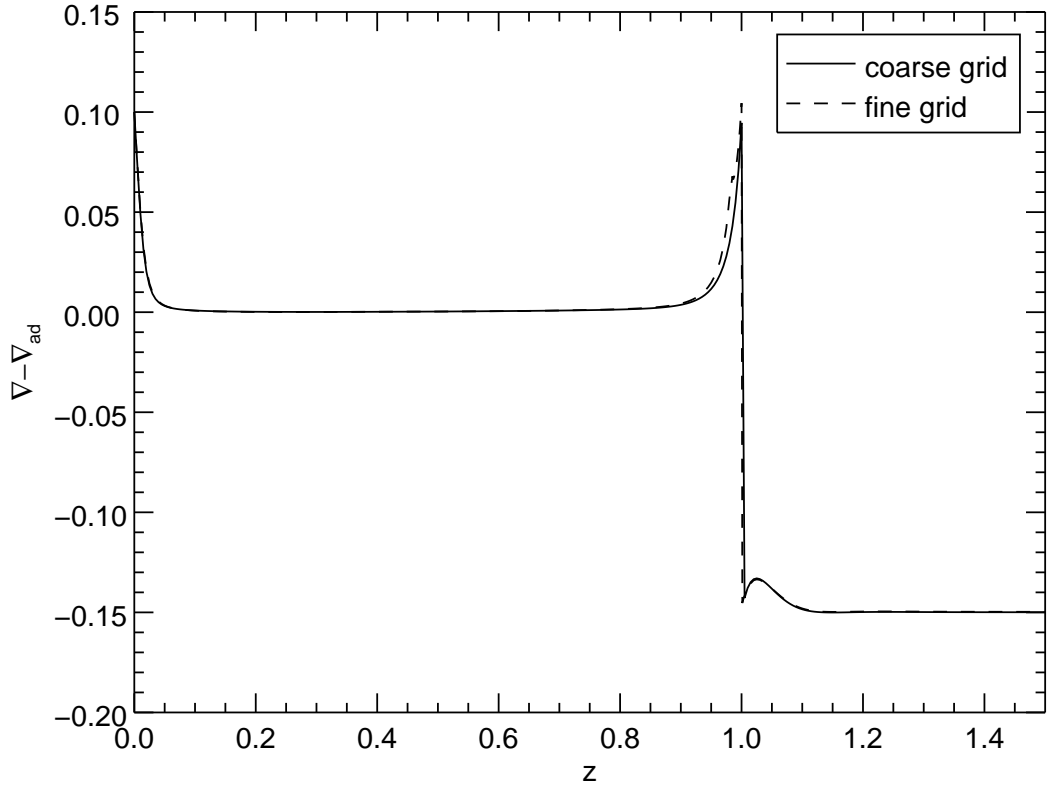}
\caption{Comparison between the coarse and fine grids. The computed case is A3. \label{fig:f8}}
\end{figure}

\appendix
The temperature gradient $\nabla$ changes abruptly near the interface. It could be a physical result or a numerical error. To justify it, we perform an additional simulation on the case A3 by adding more grid points near the interface. We insert 20 points into the region $z\in (0.98,1.02)$, thus the resolution has been improved 4 times within this region. Fig.\ref{fig:f8} compares $\nabla-\nabla_{ad}$ between the coarse and fine grid cases. We only observe slight change of $\nabla-\nabla_{ad}$ at the top of the convection zone. The shape of the structure remains the same. Thus we conclude that the abrupt change of $\nabla$ near the interface is a physical result instead of a numerical error. In our numerical experiments, the heat conductivity is initially set to be a piecewise constant function with a jump at the interface. The abrupt change of $\nabla$ near the interface is likely caused by the discontinuity of the heat conductivity. It has to be pointed out that the settings of our numerical experiments are idealized. The heat conductivity should be smooth in the real physical system. The effect of the discontinuity of the heat conductivity remains further investigation.

\acknowledgements
I thank an anonymous referee for reading the manuscript carefully and providing insightful comments. I also thank Q.-S. Zhang, Y. Li, and D.-R. Xiong for helpful discussions on their turbulent convection models. I am deeply grateful to K.L. Chan for frequent discussions on overshooting. This work was partly supported by NSFC (No. 11503097,11521101), the Science and Technology Program of Guangzhou (No. 201707010006), and the Science and Technology Development Fund, Macau SAR (No. 0045/2018/AFJ, 119/2017/A3). The simulations were performed on the supercomputers at the Purple Mountain Observatory, and the National Supercomputer Center in Guangzhou.




\bibliographystyle{aasjournal}
\bibliography{ov1}

\begin{thebibliography}{}
\expandafter\ifx\csname natexlab\endcsname\relax\def\natexlab#1{#1}\fi
\providecommand{\url}[1]{\href{#1}{#1}}
\providecommand{\dodoi}[1]{doi:~\href{http://doi.org/#1}{\nolinkurl{#1}}}
\providecommand{\doeprint}[1]{\href{http://ascl.net/#1}{\nolinkurl{http://ascl.net/#1}}}
\providecommand{\doarXiv}[1]{\href{https://arxiv.org/abs/#1}{\nolinkurl{https://arxiv.org/abs/#1}}}

\bibitem[{Arnett {et~al.}(2015)Arnett, Meakin, Viallet, Campbell, Lattanzio, \&
  Moc{\'a}k}]{arnett2015beyond}
Arnett, W.~D., Meakin, C., Viallet, M., {et~al.} 2015, The Astrophysical
  Journal, 809, 30

\bibitem[{Arnett \& Moravveji(2017)}]{arnett2017synergies}
Arnett, W.~D., \& Moravveji, E. 2017, The Astrophysical Journal Letters, 836,
  L19

\bibitem[{Browning {et~al.}(2004)Browning, Brun, \&
  Toomre}]{browning2004simulations}
Browning, M.~K., Brun, A.~S., \& Toomre, J. 2004, The Astrophysical Journal,
  601, 512

\bibitem[{Brummell {et~al.}(2002)Brummell, Clune, \&
  Toomre}]{brummell2002penetration}
Brummell, N.~H., Clune, T.~L., \& Toomre, J. 2002, The Astrophysical Journal,
  570, 825

\bibitem[{Brun {et~al.}(2005)Brun, Browning, \& Toomre}]{brun2005simulations}
Brun, A.~S., Browning, M.~K., \& Toomre, J. 2005, The Astrophysical Journal,
  629, 461

\bibitem[{Brun {et~al.}(2017)Brun, Strugarek, Varela, Matt, Augustson, Emeriau,
  DoCao, Brown, \& Toomre}]{brun2017differential}
Brun, A.~S., Strugarek, A., Varela, J., {et~al.} 2017, The Astrophysical
  Journal, 836, 192

\bibitem[{Cai(2016)}]{cai2016semi}
Cai, T. 2016, Journal of Computational Physics, 310, 342

\bibitem[{Canuto(1992)}]{canuto1992turbulent}
Canuto, V.~M. 1992, The Astrophysical Journal, 392, 218

\bibitem[{Canuto(1993)}]{canuto1993turbulent}
---. 1993, The Astrophysical Journal, 416, 331

\bibitem[{Canuto(2011)}]{canuto2011stellar}
---. 2011, Astronomy \& Astrophysics, 528, A76

\bibitem[{Canuto \& Dubovikov(1998)}]{canuto1998stellar}
Canuto, V.~M., \& Dubovikov, M. 1998, The Astrophysical Journal, 493, 834

\bibitem[{Canuto {et~al.}(2001)Canuto, Howard, Cheng, \&
  Dubovikov}]{canuto2001ocean}
Canuto, V.~M., Howard, A., Cheng, Y., \& Dubovikov, M. 2001, Journal of
  Physical Oceanography, 31, 1413

\bibitem[{Chan {et~al.}(2010)Chan, Cai, \& Singh}]{chan2010overshooting}
Chan, K.~L., Cai, T., \& Singh, H.~P. 2010, in Proceedings of the International
  Astronomical Union Symposium 271, ed. N.~Brummell, A.~Brun, M.~Miesh, \&
  Y.~Ponty (Cambridge: Cambridge Univ. Press), 317--325

\bibitem[{Claret \& Torres(2017)}]{claret2017dependence}
Claret, A., \& Torres, G. 2017, The Astrophysical Journal, 849, 18

\bibitem[{Claret \& Torres(2018)}]{claret2018dependence}
---. 2018, The Astrophysical Journal, 859, 100

\bibitem[{Cogan(1975)}]{cogan1975convective}
Cogan, B.~C. 1975, Publications of the Astronomical Society of Australia, 2,
  355

\bibitem[{Deheuvels {et~al.}(2016)Deheuvels, Brand{\~a}o, Aguirre, Ballot,
  Michel, Cunha, Lebreton, \& Appourchaux}]{deheuvels2016measuring}
Deheuvels, S., Brand{\~a}o, I., Aguirre, V.~S., {et~al.} 2016, Astronomy \&
  Astrophysics, 589, A93

\bibitem[{Deng \& Xiong(2008)}]{deng2008define}
Deng, L., \& Xiong, D.-R. 2008, Monthly Notices of the Royal Astronomical
  Society, 386, 1979

\bibitem[{Freytag {et~al.}(1996)Freytag, Ludwig, \&
  Steffen}]{freytag1996hydrodynamical}
Freytag, B., Ludwig, H.-G., \& Steffen, M. 1996, Astronomy and Astrophysics,
  313, 497

\bibitem[{Hotta(2017)}]{hotta2017solar}
Hotta, H. 2017, The Astrophysical Journal, 843, 52

\bibitem[{Hurlburt {et~al.}(1994)Hurlburt, Toomre, Massaguer, \&
  Zahn}]{hurlburt1994penetration}
Hurlburt, N.~E., Toomre, J., Massaguer, J.~M., \& Zahn, J.-P. 1994, The
  Astrophysical Journal, 421, 245

\bibitem[{K{\"a}pyl{\"a}(2018)}]{kapyla2018overshooting}
K{\"a}pyl{\"a}, P.~J. 2018, arXiv preprint arXiv:1812.07916

\bibitem[{Kitiashvili {et~al.}(2016)Kitiashvili, Kosovichev, Mansour, \&
  Wray}]{kitiashvili2016dynamics}
Kitiashvili, I.~N., Kosovichev, A.~G., Mansour, N.~N., \& Wray, A.~A. 2016, The
  Astrophysical Journal Letters, 821, L17

\bibitem[{Korre {et~al.}(2019)Korre, Garaud, \& Brummell}]{korre2019convective}
Korre, L., Garaud, P., \& Brummell, N. 2019, Monthly Notices of the Royal
  Astronomical Society, 484, 1220

\bibitem[{Kuhfu{\ss}(1986)}]{kuhfuss1986model}
Kuhfu{\ss}, R. 1986, Astronomy and Astrophysics, 160, 116

\bibitem[{Li(2012)}]{li2012k}
Li, Y. 2012, The Astrophysical Journal, 756, 37

\bibitem[{Li(2017)}]{li2017applications}
---. 2017, The Astrophysical Journal, 841, 10

\bibitem[{Li \& Yang(2007)}]{li2007testing}
Li, Y., \& Yang, J. 2007, Monthly Notices of the Royal Astronomical Society,
  375, 388

\bibitem[{Maeder(1975)}]{maeder1975stellar}
Maeder, A. 1975, Astronomy and Astrophysics, 40, 303

\bibitem[{Moravveji {et~al.}(2015)Moravveji, Aerts, P{\'a}pics, Triana, \&
  Vandoren}]{moravveji2015tight}
Moravveji, E., Aerts, C., P{\'a}pics, P.~I., Triana, S.~A., \& Vandoren, B.
  2015, Astronomy \& Astrophysics, 580, A27

\bibitem[{Moravveji {et~al.}(2016)Moravveji, Townsend, Aerts, \&
  Mathis}]{moravveji2016sub}
Moravveji, E., Townsend, R.~H., Aerts, C., \& Mathis, S. 2016, The
  Astrophysical Journal, 823, 130

\bibitem[{Rempel(2004)}]{rempel2004overshoot}
Rempel, M. 2004, The Astrophysical Journal, 607, 1046

\bibitem[{Renzini(1987)}]{renzini1987some}
Renzini, A. 1987, Astronomy and Astrophysics, 188, 49

\bibitem[{Roxburgh(1965)}]{roxburgh1965note}
Roxburgh, I.~W. 1965, Monthly Notices of the Royal Astronomical Society, 130,
  223

\bibitem[{Roxburgh(1978)}]{roxburgh1978convection}
---. 1978, Astronomy and Astrophysics, 65, 281

\bibitem[{Saikia {et~al.}(2000)Saikia, Singh, Chan, Roxburgh, \&
  Srivastava}]{saikia2000examination}
Saikia, E., Singh, H.~P., Chan, K., Roxburgh, I., \& Srivastava, M. 2000, The
  Astrophysical Journal, 529, 402

\bibitem[{Saslaw \& Schwarzschild(1965)}]{saslaw1965overshooting}
Saslaw, W., \& Schwarzschild, M. 1965, The Astrophysical Journal, 142, 1468

\bibitem[{Schmitt {et~al.}(1984)Schmitt, Rosner, \&
  Bohn}]{schmitt1984overshoot}
Schmitt, J., Rosner, R., \& Bohn, H. 1984, The Astrophysical Journal, 282, 316

\bibitem[{Shaviv \& Salpeter(1973)}]{shaviv1973convective}
Shaviv, G., \& Salpeter, E.~E. 1973, The Astrophysical Journal, 184, 191

\bibitem[{Singh {et~al.}(1994)Singh, Roxburgh, \& Chan}]{singh1994three}
Singh, H.~P., Roxburgh, I.~W., \& Chan, K.~L. 1994, Astronomy and Astrophysics,
  281, L73

\bibitem[{Singh {et~al.}(1995)Singh, Roxburgh, \& Chan}]{singh1995three}
---. 1995, Astronomy and Astrophysics, 295, 703

\bibitem[{Singh {et~al.}(1998)Singh, Roxburgh, \& Chan}]{singh1998study}
---. 1998, Astronomy and Astrophysics, 340, 178

\bibitem[{Stancliffe {et~al.}(2015)Stancliffe, Fossati, Passy, \&
  Schneider}]{stancliffe2015confronting}
Stancliffe, R.~J., Fossati, L., Passy, J.-C., \& Schneider, F. 2015, Astronomy
  \& Astrophysics, 575, A117

\bibitem[{Viallet {et~al.}(2015)Viallet, Meakin, Prat, \&
  Arnett}]{viallet2015toward}
Viallet, M., Meakin, C., Prat, V., \& Arnett, D. 2015, Astronomy \&
  Astrophysics, 580, A61

\bibitem[{Xiong(1978)}]{xiong1978stochastic}
Xiong, D.-R. 1978, Chinese Astronomy, 2, 118

\bibitem[{Xiong(1981)}]{xiong1981statistical}
---. 1981, Scientia Sinica, 24, 1406

\bibitem[{Xiong(1986)}]{xiong1986evolution}
---. 1986, Astronomy and Astrophysics, 167, 239

\bibitem[{Xiong(1989)}]{xiong1989radiation}
---. 1989, Astronomy and Astrophysics, 209, 126

\bibitem[{Xiong {et~al.}(1997)Xiong, Cheng, \& Deng}]{xiong1997nonlocal}
Xiong, D.-R., Cheng, Q., \& Deng, L. 1997, The Astrophysical Journal Supplement
  Series, 108, 529

\bibitem[{Xiong \& Deng(2001)}]{xiong2001structure}
Xiong, D.-R., \& Deng, L. 2001, Monthly Notices of the Royal Astronomical
  Society, 327, 1137

\bibitem[{Zahn(1991)}]{zahn1991convective}
Zahn, J.-P. 1991, Astronomy and Astrophysics, 252, 179

\bibitem[{Zhang(2012)}]{zhang2012testing}
Zhang, Q.-S. 2012, The Astrophysical Journal, 761, 153

\bibitem[{Zhang(2013)}]{zhang2013convective}
---. 2013, The Astrophysical Journal Supplement Series, 205, 18

\bibitem[{Zhang(2016)}]{zhang2016simple}
---. 2016, The Astrophysical Journal, 818, 146

\bibitem[{Zhang \& Li(2012)}]{zhang2012turbulent}
Zhang, Q.-S., \& Li, Y. 2012, The Astrophysical Journal, 750, 11

\end{thebibliography}



\end{document}